\documentclass[numberedappendix]{emulateapj}

\usepackage{epsfig, natbib}
\tightenlines

\countdef\decade=200
\advance\decade by \year
\countdef\hours=201	
\advance\hours by \time
\divide\hours by 60
\countdef\mins=202
\advance\mins by \hours
\multiply\mins by 60
\multiply\hours by 100
\countdef\miltime=203
\advance\miltime by \hours
\advance\miltime by \time
\advance\miltime by -\mins

\def\mathbi#1{\textbf{\em #1}}

\newcommand{\gtsim}{\protect\raisebox{-0.5ex}{$\:\stackrel{\textstyle >}
        {\sim}\:$}}

\newcommand{\msun}{M_{\odot}}

\newcommand{\tff}{t_{\rm ff}}
\newcommand{\vecv}{\mathbi{v}}
\newcommand{\vecx}{\mathbi{x}}
\newcommand{\vecp}{\mathbi{p}}

\newcommand{\jcam}{J.\ Comp.\ App.\ Math.}
\newcommand{\jcompphys}{JCP}
\newcommand{\orion}{\textsc{orion}~}
\newcommand{\Orion}{\textsc{Orion}~}
\newcommand{\na}{New Astron.}

\usepackage{color}
\newcommand{\red}[1]{#1}

\begin{document}

\title{Radiation-Hydrodynamic Simulations of the Formation of Orion-Like Star Clusters\\
II. The Initial Mass Function from Winds, Turbulence, and Radiation}

\slugcomment{Accepted for publication in The Astrophysical Journal}

\shorttitle{Radiation-Hydrodynamic Simulations of Cluster Formation II.}
\shortauthors{Krumholz et al.}

\author{
        Mark R. Krumholz\altaffilmark{1},
        Richard I. Klein\altaffilmark{2, 3}, and
        Christopher F. McKee\altaffilmark{3,4}}

\altaffiltext{1}{Department of Astronomy and Astrophysics,
         University of California, Santa Cruz, CA 95064;
         krumholz@ucolick.org}
\altaffiltext{2}{Lawrence Livermore National Laboratory, P.O. Box 808, L-23, Livermore, CA 94550}
\altaffiltext{3}{Department of Astronomy and Astrophysics, University of California, Berkeley,
Berkeley, CA 94720}
\altaffiltext{4}{Department of Physics, University of California, Berkeley,
Berkeley, CA 94720}

\begin{abstract}
We report a series of simulations of the formation of a star cluster similar to the Orion Nebula Cluster (ONC), including both radiative transfer and protostellar outflows, and starting from both smooth and self-consistently turbulent initial conditions. 
\red{Each simulation forms $>150$}
stars and brown dwarfs, yielding a stellar mass distribution that
\red{ranges}
from $< 0.1$ $\msun$ to $> 10$ $\msun$. We show that a simulation that begins with self-consistently turbulent density and velocity fields embedded in a larger turbulent volume, and that includes protostellar outflows, produces \red{an initial mass function (IMF) that is consistent both with that of the ONC and the Galactic field, at least within the statistical power provided by the number of stars formed in our simulations}.
This is the first simulation published to date that reproduces the observed IMF in a cluster large enough to contain massive stars, and where the peak of the mass function is determined by a fully self-consistent calculation of gas thermodynamics rather than a hand-imposed equation of state. This simulation also produces a star formation rate that, while still somewhat too high, is much closer to observed values than 
\red{if we omit either the larger turbulent volume or the outflows}.
Moreover, we show that the combination of outflows, \red{self-consistently turbulent initial conditions}, and turbulence continually fed by motions on scales larger than that of the protocluster yields an IMF that is \red{in agreement with observations and} invariant with time, resolving the ``overheating" problem in which simulations without these features have an IMF peak that shifts to progressively higher masses over time as more and more of the gas is heated, inconsistent with the observed invariance of the IMF. The simulation that matches the observed IMF also \red{qualitatively} reproduces the observed trend of stellar multiplicity strongly increasing with mass. We show that this simulation produces massive stars from distinct massive cores whose properties are consistent with those of observed massive cores. However, the stars formed in these cores also undergo dynamical interactions as they accrete that naturally produce Trapezium-like hierarchical multiple systems of massive stars.
\end{abstract}


\section{Introduction}
\label{sec:intro}

The origin of the the stellar initial mass function (IMF) is a classic problem in astrophysics. Since the IMF is most easily measured in young star clusters, and appears to be essentially the same in such clusters and in the field \citep[e.g.][]{bastian10a}, this problem is closely linked to the problem of how star clusters form. There have been numerous theoretical attacks on these twin problems (see the review by \citealt{mckee07a}), but a major breakthrough in the past few years has been the realization that the answer is tightly linked to the question of gas thermodynamics. An isothermal gas, even a magnetized one, has no characteristic mass scale \citep{mckee10b, krumholz11e}. This implies that the problem of the origin of the IMF, which is observed to be invariant in both its shape and its characteristic scale, is a separable one.

Models that describe the behavior of a isothermal gas, such as those based on turbulent fragmentation \citep[e.g.][]{padoan02a, hennebelle08b} or competitive accretion \citep[e.g.][]{bonnell01a, bonnell01b}, can predict a shape for the IMF, but in order to determine a characteristic scale must either appeal to additional physics or must define a fiducial ``cloud", whose mean density or other properties (e.g.\ the normalization of its linewidth-size relation) then determines the location of the IMF peak. In the latter approach, however, it is not clear on what scale one should measure cloud properties: an entire GMC, with a mean density $n\sim 10^2$ cm$^{-3}$ obeying the \citet{larson81} linewidth-size relation, a massive clump with a mean density $n\sim 10^5$ cm$^{-3}$ and a linewidth far above the \citeauthor{larson81} value \citep[e.g.][]{shirley03a}, or some other scale? Different choices yield wildly varying characteristic masses. Moreover, cloud properties vary in sufficiently extreme galactic environments, for example showing different linewidth-size relations \citep[e.g.][]{rosolowsky05a}. Despite this variation, however, there is no evidence for a corresponding variation in the IMF. These problems strongly suggest that the origin of the IMF peak cannot be found in the physics of isothermal gas. Instead, models that seek to explain any characteristic mass scale in the IMF must appeal to departures from isothermality \citep{rees76a, low76a, spaans00a, larson05a, krumholz11e}.

Simulations of star formation mirror these trends depending on the physics they include. Isothermal simulations always produce a characteristic stellar mass that is determined by the initial conditions or the numerical resolution \citep[e.g.][]{martel06a}, and can always be rescaled to produce an arbitrary stellar mass scale. In contrast, those that include non-isothermality produce characteristic mass scales that are determined by the mechanism that causes them to depart from isothermality, whether it be an imposed equation of state \citep[e.g.][]{bate05a, jappsen05a} or the inclusion of radiative transfer, either without \citep{bate09a, bate12a} or with \citep{krumholz07a, krumholz10a, krumholz11c, offner09a, urban10a} the further step of including stellar radiation. Since comparisons between approximate equations of state and radiative transfer calculations show that the former offer only an extremely poor approximation, progress toward an understanding of the IMF's characteristic peak therefore requires radiation-hydrodynamic simulations.

The radiation-hydrodynamic simulations that have been conducted thus far have demonstrated several promising features. First, radiation feedback suppresses the formation of brown dwarfs, reproducing the observed turn-down in the IMF at low masses \citep{bate09a, offner09a}. Second, simulations including radiation feedback are able to suppress fragmentation in very dense regions, allowing the formation of massive stars when the conditions are approach those seen in real regions of massive star formation \citep{krumholz07a, krumholz10a}. Third, radiative simulations produce an IMF that does not vary with the properties of the star-forming cloud in low mass, low-density environments \citep{bate09a, bate12a}, nor with the gas metallicity \citep{myers11a}.

While these results are encouraging, these simulation efforts have for the most part been limited either to single massive cores, or to clouds of low density and/or low mass. For example, \citet{bate12a} simulates a cloud of mean volume density $3\times 10^4$ cm$^{-3}$ and column density $0.2$ g cm$^{-2}$, forming $\sim 80$ $\msun$ of stars, none larger than $\sim 3$ $\msun$. \citet{peters10a} do form massive stars, but from a cloud with a mean density of $10^3$ cm$^{-3}$ and a column density of $0.026$ g cm$^{-2}$, far below the column density at which radiative effects become important \citep{krumholz08a, krumholz10a} -- so low, in fact, as to be optically thin in the near-infrared. In contrast, the mean mass and radius of the star-forming regions studied by \citet{faundez04a} implies a volume density $>10^5$ cm$^{-3}$, a mass of $\sim 5000$ $\msun$, and a column density of 2 g cm$^{-2}$, such that multiple massive stars would be expected, and their radiation would be trapped effectively by the cloud's high optical depth. Similar Galaxy-wide surveys by \citet{shirley03a} and \citet{fontani05a} that target regions of active star formation produce comparable properties. Indeed, the observed cluster mass function is $dN/dM \propto M^{-2}$ \citep{lada03a, fall09a, chandar10a}, implying that a majority of stars form in clusters larger than $1000$ $\msun$ in mass, large enough to possess O stars. The ONC, therefore, is a far more typical star-forming environment than most of the regions explored with radiation-hydrodynamic simulations thus far. 

In \citet[hereafter Paper I]{krumholz11c} we reported the first radiation-hydrodynamic simulations to probe this more typical regime of star formation; that calculation followed the collapse of a $1000$ $\msun$ cloud with a column density of 1 g cm$^{-2}$, leading to the production of $>500$ $\msun$ worth of stars, with an IMF extending from $\sim 0.05$ $\msun$ brown dwarfs to $\sim 30$ $\msun$ O stars. This calculation identified a problem. Radiative suppression of fragmentation, which seems necessary to explain the invariant peak in the IMF and avoid overproduction of brown dwarfs, became too efficient. As the calculation proceeded, the cloud underwent a global collapse, leading to extremely high star formation rates and accretion luminosities. As a result, the gas heated up to the point where further star formation was suppressed. The net result was an IMF that was not invariant, but instead had a peak that moved to systematically higher masses as the calculation proceeded. At early times there were too few massive stars, and at late times too many. Since there is no plausible mechanism to guarantee that all star-forming clouds would stop producing stars at the same point in this evolution, this result was inconsistent with the observed universality of the IMF.

In Paper I, we conjectured that the problem could be resolved by lowering the star formation rate per free-fall time, which would in turn lower the accretion luminosity. Such a change is required by observations even in the absence of the problems rapid star formation creates in the IMF, because observed star formation rates per free-fall time are always a few percent across a very wide range of star-forming environments \citep{krumholz07e, evans09a, krumholz12a}. In this paper we test that conjecture by performing additional simulations of the formation of ONC-like star clusters, with two extra pieces of physics that should lower the star formation rate per free-fall time. First, rather than simulating an isolated star-forming clump as in Paper I, we embed our initial clump in a larger volume of turbulent gas, and we initialize the simulations such that our clump has self-consistently generated turbulent density and velocity structure. Second, we include protostellar outflows. A number of authors have shown that such outflows can inject significant energy into a star-forming cloud, driving its turbulence and lowering its star formation rate \citep{li06b, nakamura07a, matzner07a, wang10a, cunningham11a}. Ideally, a third piece of physics should also be included: magnetic fields. These both lower the star formation rate by themselves \red{\citep[e.g.][]{price09a}}, and also enhance the effectiveness of protostellar outflows \citep{wang10a}. We plan to do so in future work.

The remainder of this paper is organized as follows. In Section \ref{sec:methods} we describe our numerical methods and simulation setup. In Section \ref{sec:results} we report the simulation results, and finally we discuss their implications and draw conclusions in Section \ref{sec:discussion}.

\section{Numerical Methods}
\label{sec:methods}

We simulate star-forming clouds using the \orion code, which includes radiative transfer \citep{howell03a, krumholz07b, shestakov08a}, hydrodynamics \citep{klein99a}, self-gravity \citep{truelove98a}, accreting sink particles \citep{krumholz04a}, and a model for protostellar evolution and feedback, including stellar radiation and outflows \citep{offner09a, cunningham11a}. Here we briefly summarize the equations we solve, the code itself, and the initial conditions for the simulations. For the first two of these topics, we refer the reader to Paper I for more details, since the physics included and the numerical methods are identical except where specified below.

\subsection{Equations and Algorithms}

\orion solves the equations of gravito-radiation-hydrodynamics in the two-temperature, mixed-frame flux-limited diffusion approximation. These equations are \citep{krumholz07b}
\begin{eqnarray}
\frac{\partial}{\partial t}\rho & = & - \nabla\cdot(\rho\vecv) - \sum_i \dot{M}_{a,i} W_a(\vecx-\vecx_i) 
\nonumber \\
& & {} + \sum_i \dot{M}_{w,i} W_w (\vecx-\vecx_i) 
\label{eq:masscons}
\\
\frac{\partial}{\partial t}(\rho \vecv) & = & -\nabla\cdot(\rho \vecv\vecv) - \nabla P - \rho \nabla \phi - \lambda \nabla E
\nonumber \\
& & {} - \sum_i \dot{\vecp}_{a,i} W_a(\vecx-\vecx_i) +  \sum_i \dot{\vecp}_{w,i} W_w(\vecx-\vecx_i) 
\label{momcons}
\\
\frac{\partial}{\partial t}(\rho e) & = & -\nabla \cdot [(\rho e+P)\vecv] - \rho \vecv \cdot \nabla \phi - \kappa_{\rm 0P} \rho (4 \pi B - c E) 
\nonumber \\
& & {} + \lambda\left(2 \frac{\kappa_{\rm 0P}}{\kappa_{\rm 0R}} - 1\right) \vecv \cdot \nabla E - \left(\frac{\rho}{m_p}\right)^2 \Lambda(T_g) \\
& & {} - \sum_i \dot{\mathcal{E}}_{a,i} W_a(\vecx - \vecx_i) + \sum_i \dot{\mathcal{E}}_{w,i} W_w(\vecx - \vecx_i) 
\label{econsgas}
\\
\frac{\partial}{\partial t}E & = & \nabla \cdot \left(\frac{c\lambda}{\kappa_{\rm 0R} \rho} \nabla E\right) + \kappa_{\rm 0P} \rho (4 \pi B - c E) 
\nonumber \\
& & {} - \lambda \left(2\frac{\kappa_{\rm 0P}}{\kappa_{\rm 0R}} - 1\right) \vecv\cdot \nabla E - \nabla \cdot \left(\frac{3 - R_2}{2} \vecv E\right)
\nonumber \\
& & {}
 +  \left(\frac{\rho}{m_p}\right)^2 \Lambda(T_g) + \sum_i L_i W(\vecx - \vecx_i)
\label{econsrad} 
\\
\label{starmass}
\frac{d}{dt} M_i &= & \dot{M}_i \\
\label{starpos}
\frac{d}{dt} \vecx_i & = & \frac{\vecp_i}{M_i} \\
\label{starmom}
\frac{d}{dt} \vecp_i & = & -M_i \nabla \phi + \dot{\vecp}_i
\\
\label{eq:poisson}
\nabla^2\phi & = &4\pi G \left[ \rho + \sum_i M_i \delta(\vecx-\vecx_i)\right].
\end{eqnarray}
In these equations, $\rho$, $\vecv$, $P$, and $e$ are the density, velocity, pressure, and total (thermal plus kinetic) energy density of the gas, $E$ is the energy density of the radiation, $\phi$ is the gravitational potential, $\kappa_{\rm 0P}$ and $\kappa_{\rm 0R}$ are the Planck and Rosseland mean opacities of the dust-plus-gas fluid, $\lambda$ is the flux-limiter, $\Lambda$ is the rate of non-dust cooling (via line and continuum processes in gas at temperatures $\gtsim 10^3$ K where the dust sublimes), and $m_p$ is the proton mass. For more information on the flux-limiter, hot gas cooling rate, and choice of dust opacities, we refer the reader to Paper I.

Terms subscripted by $i$ refer to stars; $\vecx_i$, $M_i$, and $\vecp_i$ are the position, mass, and momentum of the $i$th star, $\dot{M}_i$, $\dot{\vecp}_i$, and $\dot{\mathcal{E}}_i$ are the rate at which those stars add or remove mass, momentum, and energy from the gas, $L_i$  is the luminosity of star $i$, and $W_i$ is the weighting kernel that spreads the stellar interaction over some number of computational cells. The equations we solve here differ from those in Paper I in that, in addition to accretion (the terms subscripted with $a$), we also include protostellar winds (the terms subscripted with $w$). Stars accrete gas from the computational grid following the sink particle method of \citet{krumholz04a}, and each sink particle is linked to a protostellar evolution code that computes the instantaneous stellar radius and luminosity based on the star's accretion history, following the method described in the Appendix of \citet{offner09a}. In addition, during each time step, each star returns a portion of the mass it accretes to the grid in the form of a collimated protostellar wind. For details of the numerical implementation, see \citet{cunningham11a}. Our wind parameters are the same as in that paper, i.e.~each star ejects a fraction $f_w/(1+f_w)=0.21$ of the gas it accretes (so $f_w = 0.27$), this material is launched with a velocity $f_v = 1/3$ that of the Keplerian speed at the stellar surface, and the wind gas has a temperature $10^4$ K at launch. It is collimated along the axis defined by the stellar angular momentum vector.

\Orion solves Equations (\ref{eq:masscons}) -- (\ref{eq:poisson}) within an overall adaptive mesh refinement (AMR) structure, in which the entire domain is discretized onto a coarse grid of size $N_0$ cells on a side, denoted level 0. Sub-regions within the domain are then covered by progressively finer grids. The grid on level $\ell$ has a resolution a factor of $2^\ell$ better than that of the coarse grid, and evolves with a time step a factor of $2^\ell$ smaller. These grids are automatically added and removed on the fly as the calculation proceeds, based on user-specified criteria, up to some pre-specified maximum level $L$.

\subsection{Simulation Setup}

\begin{deluxetable*}{ccccccccccc}
\tablecaption{Simulation Parameters\label{tab:runparam}}
\tablehead{
\colhead{Name} &
\colhead{Winds?} &
\colhead{$M_c$} &
\colhead{$\ell_c$ or $R_c$} &
\colhead{$\sigma_c$} &
\colhead{$\langle\rho\rangle_M$} &
\colhead{$\tff$} &
\colhead{$\ell_{\rm box}$} &
\colhead{$N_0$} &
\colhead{$L$} &
\colhead{$\Delta x_L$}
\\
\colhead{} &
\colhead{} &
\colhead{($\msun$)} &
\colhead{(pc)} &
\colhead{(km s$^{-1}$)} &
\colhead{(g cm$^{-3}$)} &
\colhead{(kyr)} &
\colhead{(pc)} &
\colhead{} &
\colhead{} &
\colhead{(AU)}
}

\startdata
SmNW & No & 1000 & 0.26 & 2.9 & $1.4\times 10^{-18}$ & 56 & 1.9 & 256 & 5 & 49 \\
TuNW & No & 1000 & 0.46 & 1.4 & $8.6\times 10^{-18}$ & 23 & 0.46 & 256 & 4 & 23 \\
TuW & Yes & 1000 & 0.46 & 1.4 & $8.6\times 10^{-18}$ & 23 & 0.46 & 256 & 4 & 23 \\
\enddata
\tablecomments{Col.\ 3: cloud mass. Col.\ 4: cloud radius (for run SmNW) or box size (for runs TuNW and TuW). Col.\ 6: mass weighted-mean density at time $t=0$. Col.\ 7: free-fall time computed using $\langle\rho\rangle_M$. Col.\ 8: size of computational box. Col.\ 9: number of cells per linear dimension on the coarsest AMR level. Col.\ 10: finest AMR level. Col.\ 11: grid resolution on the finest AMR level.
}
\end{deluxetable*}

We compare three different simulations, which we refer to as smooth, no wind (SmNW), turbulent, no wind (TuNW), and turbulent, with winds (TuW); in terms of physics these differ in that the NW simulations have protostellar outflows disabled. We summarize this and other properties of the simulations in Table \ref{tab:runparam}. All simulations consist of a mass $M_c = 1000$ $\msun$ of gas with a mean surface density $\Sigma_c = 1$ g cm$^{-2}$ (arranged as described below). Throughout the cloud we set the gas temperature and the radiation energy density to $T_g = 10$ K and $E = a T_g^4 = 7.56\times 10^{-11}$ erg cm$^{-3}$, respectively.

For run SmNW, we use a setup identical to run HR from Paper I (though here we have continued the simulation further in time than we described in that paper), so we only briefly discuss its properties here, and refer readers to Paper I for a fuller description. In run SmNW, the initial gas distribution is a sphere with a radius $R_c = 0.26$ pc.  The density distribution is smooth, and consists of a central core of uniform density that extends to half the cloud's radius, surrounded by an outer region within which the density falls off with radius as $r^{-1.5}$, as suggested by observations of massive clumps \citep[e.g.][]{sridharan05a, beuther06a}. The gas is given an initial turbulent velocity field with a dispersion of $\sigma_c = 2.9$ km s$^{-1}$ (one-dimensional), corresponding to an initial virial ratio $\alpha = 5 \sigma_c^2 R_c / G M_c = 2.5$. The velocity power spectrum is $P(k) \propto k^{-2}$, drawn without imposing any bias in favor of solenoidal or compressive modes following the procedure of \citet{dubinski95a}. Outside the sphere of gas we place a zero-opacity ambient medium with a temperature 100 times larger and a density 100 times smaller than that of the gas at the sphere's edge. \red{We emphasize that, because the density gradient in the gas only extends to half the initial radius, the overall center to edge density contrast is only a factor of $2.8$, substantially less than that induced by the turbulent shocks. Thus this initial condition is quite similar to that adopted by other authors who have simulated isolated clouds, e.g.~\citet{bonnell03a} and \citet{bate12a}.}\footnote{An additional difference between our setup and that of \citet{bonnell03a} and \citet{bate12a} is that we place an ambient medium outside our cloud that is in thermal pressure balance with the material at the cloud edge, while the smoothed particle hydrodynamics (SPH) simulations of \citeauthor{bonnell03a}~and \citeauthor{bate12a} have a vacuum outside their clouds. However, this difference is almost certainly negligible. The thermal pressure of our ambient medium is set equal to the thermal pressure of the cloud, which is smaller than either the ram pressure or the self-gravitational weight of the cloud by a factor of $\sim 100$. Thus the extra pressure provided by the external medium will enhance the collapse that would occur due to gravity alone by only $\sim 1\%$. Even this is likely an overestimate of the difference between the two simulation methods, because, while formally the SPH simulations have vacuum outside their clouds, SPH creates an artificial surface tension at density discontinuities \citep{price08a}, and this will act very much like a confining external pressure. Our Eulerian simulation method does not suffer from this problem.}

\begin{figure}
\epsscale{1.1}
\plotone{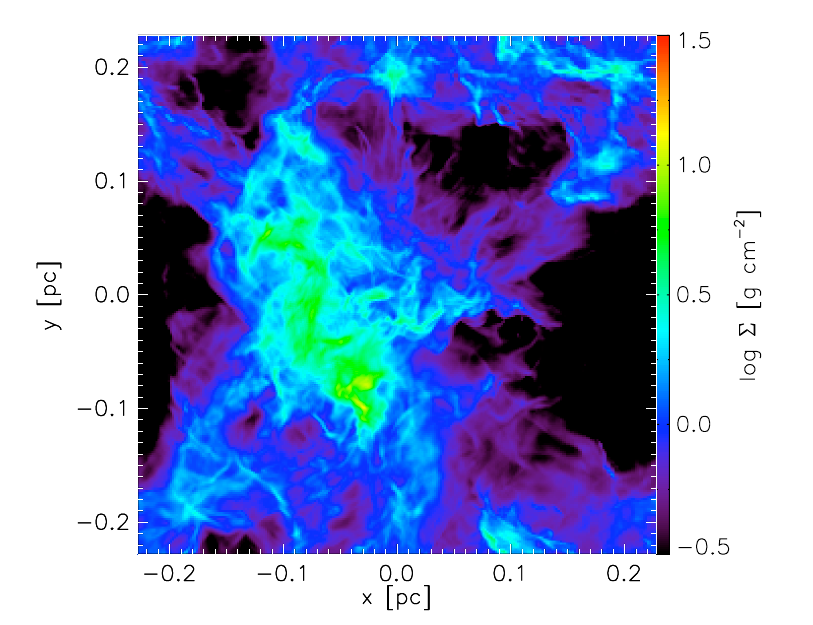}
\caption{
\label{fig:turbic}
Column density distribution in the turbulent initial conditions used for runs TuNW and TuW.
}
\end{figure}

In runs TuNW and TuW we initialize so that, unlike in run SmNW, both the initial density and velocity fields are self-consistently turbulent. We set up a periodic domain of length $\ell_c=0.46$ pc on a side, so that $\Sigma_c = 1$ g cm$^{-2}$ averaged over the box. To initialize the simulation, we impose the same turbulent velocity field as in run SmNW, scaled to a velocity dispersion $\sigma_c = 1.4$ km s$^{-1}$, corresponding to $\alpha = 1/2$ if we use $\ell_c/2$ in place of $R_c$. Although this means the gas is less turbulent initially than in run SmNW, as we see below, damping of the turbulence in run SmNW brings the $\alpha$ values closer together as the runs progress. To produce a density field consist with this velocity field, we drive the turbulence and allow the simulation to evolve for two crossing times. During this period we turn off both gravity and radiation, and we hold the gas isothermal at a temperature $T_g = 10$ K by setting the gas ratio of specific heats to $\gamma = 1.0001$; since, in the absence of stellar sources, molecular cloud gas is close to isothermal, this should be a very good approximation, and ignoring radiation during this setup phase significantly reduces the computational cost. During this setup phase we also fix the computational resolution at $512^3$ cells, with no further refinement. At the end of two crossing times we \red{turn off driving,} change the gas ratio of specific heats to $\gamma = 5/3$, turn on gravity and radiation, and return to our normal refinement criteria (see below). This state represents the initial condition for runs TuNW and TuW. Note that, since the turbulence is driven mostly on large scales, the result of this procedure is essentially a single, dense, turbulent cloud, surrounded by lower density turbulent material; we show this state in Figure \ref{fig:turbic}. This clump is therefore analogous to the isolated one in run SmNW, but is surrounded by a realistic turbulent environment rather than an artificial hot ambient medium.

In all simulations the refinement criteria used to add higher resolution grids are the same. Specifically, we add resolution in any cell that satisfies one of the following three conditions: (1) the density in the cell exceeds the local Jeans density \citep{truelove97a}, $\rho_J = J^2 \pi c_s^2 / G \Delta x_l^2$, where $J = 1/4$ is the Jeans number, $c_s = \sqrt{k_B T/\mu}$ is the isothermal sound speed, and $\Delta x_l$ is the grid spacing on AMR level $l$; (2) the radiation energy gradient is sharp enough so that $|\nabla E|/E > 0.15/\Delta x_l$ (although we sometimes temporarily reduce the coefficient below 0.15 for stability reasons in the TuNW and TuW runs); (3) the cell is within a distance of $16\Delta x_l$ of any star particle. We refine to a maximum resolution of 49 AU ($L = 5$) in run SmNW, and 23 AU ($L=4$) in runs TuNW and TuW. Finally, we note that, while the hydrodynamic and gravitational boundary conditions are necessarily different in the smooth and turbulent runs, the great majority of the star formation in the turbulent runs occurs in subregions much smaller than the entire computational volume, and thus the periodic boundary conditions have minimal impact. We also impose Marshak boundary conditions on the radiation in runs TuNW and TuW in order to let radiation escape the computational volume (c.f.~\citealt{offner09a}).

\section{Results}
\label{sec:results}

Before examining the results of our simulations, we first mention two subtleties in the analysis that apply to the remainder of this discussion. First, since we are comparing runs with different initial conditions, it is important to normalize the times so that differences between the runs reflect the underlying physical behavior, and not simply that the dynamical time is different in different cases. Moreover, in the runs with turbulent initial conditions, the strong initial turbulence guarantees that the majority of the mass is compressed into structures that are significantly denser than the volume-averaged density. Given these considerations, the most natural approach, which we adopt, is to measure times in units of the free-fall time $\tff=\sqrt{3\pi/32 G \rho}$ evaluated at a density equal to the initial mass-weighted density $\langle \rho\rangle_M$, since this is the dynamical time appropriate to the bulk of the matter. This approach also has the advantage that it is the most natural basis for observational comparison, since an observation would detect the bulk of the mass, and would be sensitive to the typical density at which this mass resides. For this reason, in what follows whenever we refer to times, we normalize to $\tff$ defined in this manner. We report this quantity in Table \ref{tab:runparam}.\footnote{Note that in Paper I we instead used the volume weighted mean density to compute the free-fall time for run SmNW; however, because the initial density field is very smooth, the difference between volume- and mass-weighted mean density free-fall times for this run is only $\sim 20\%$.} The second subtlety is that, as in Paper I, we only regard stars as collapsed objects once their mass exceeds $0.05$ $\msun$, based on one-dimensional calculations of the mass at which second collapse to stellar densities occurs \citep{masunaga00a}. We allow smaller objects to merge with one another and with more massive stars. We therefore restrict our analysis to objects larger than this mass.

\subsection{Overall Evolution and Morphology}

\begin{figure}
\plotone{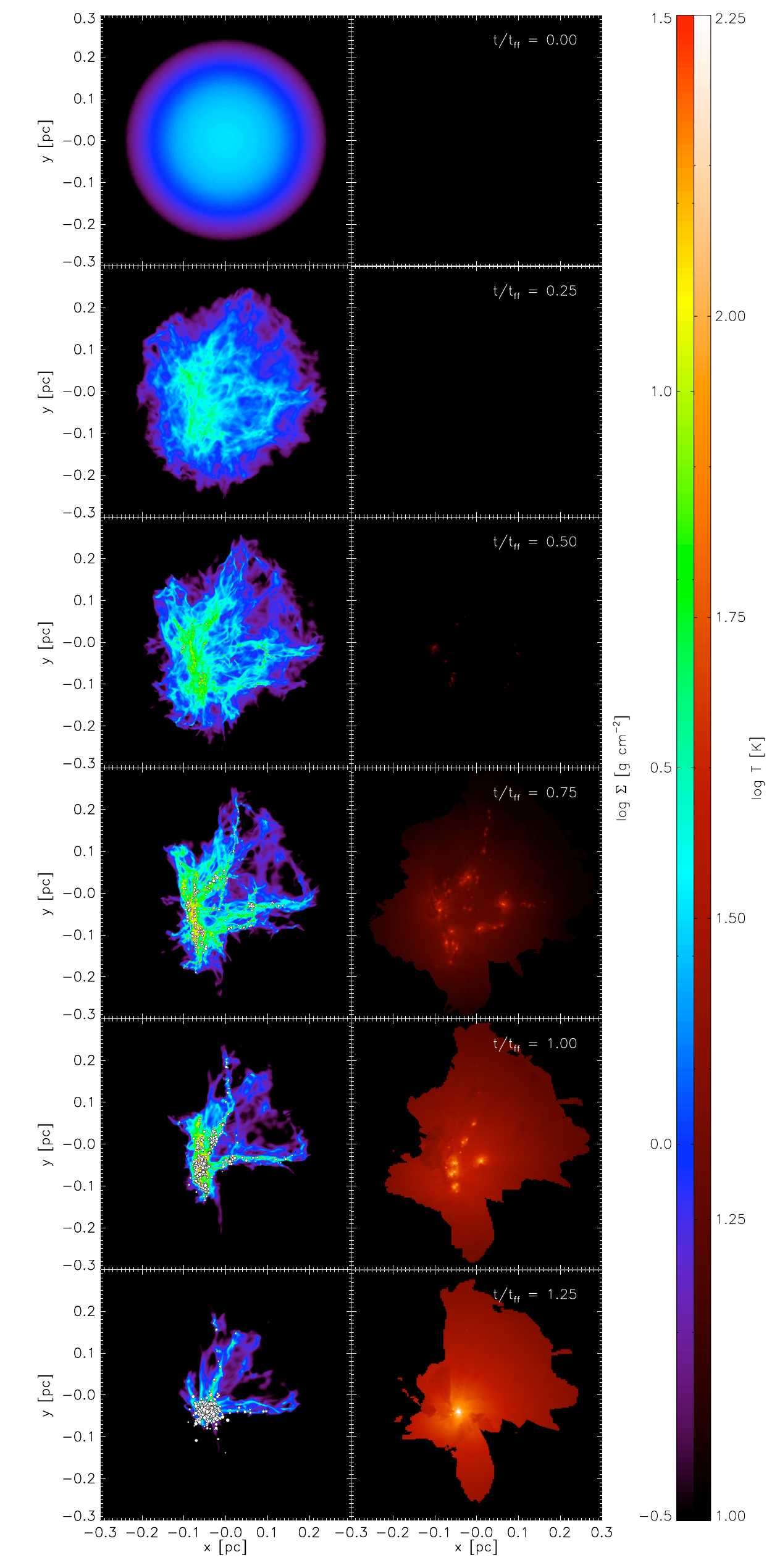}
\caption{
\label{fig:snapshots_smnw}
Column density (left) and density-weighted mean temperature (right) in run SmNW. The times of each pair of images are indicated in the right column, running from $t/\tff = 0$ to $1.25$ in steps of 0.25. In the column density plot, white circles indicate the positions of star particles, with the size of the circle indicating the mass of the star. In the right column, the temperature shown is the radiation temperature $T_r$, defined implicitly by $E=aT_r^4$. We show this rather than the gas temperature because the gas and radiation temperatures are nearly equal everywhere in the cloud, except in the hot ambient medium outside the cloud in run SmNW, and in material ejected by protostellar outflows in run TuW. Using the radiation temperature provides a convenient means to filter this contribution.
}
\end{figure}

\begin{figure}
\plotone{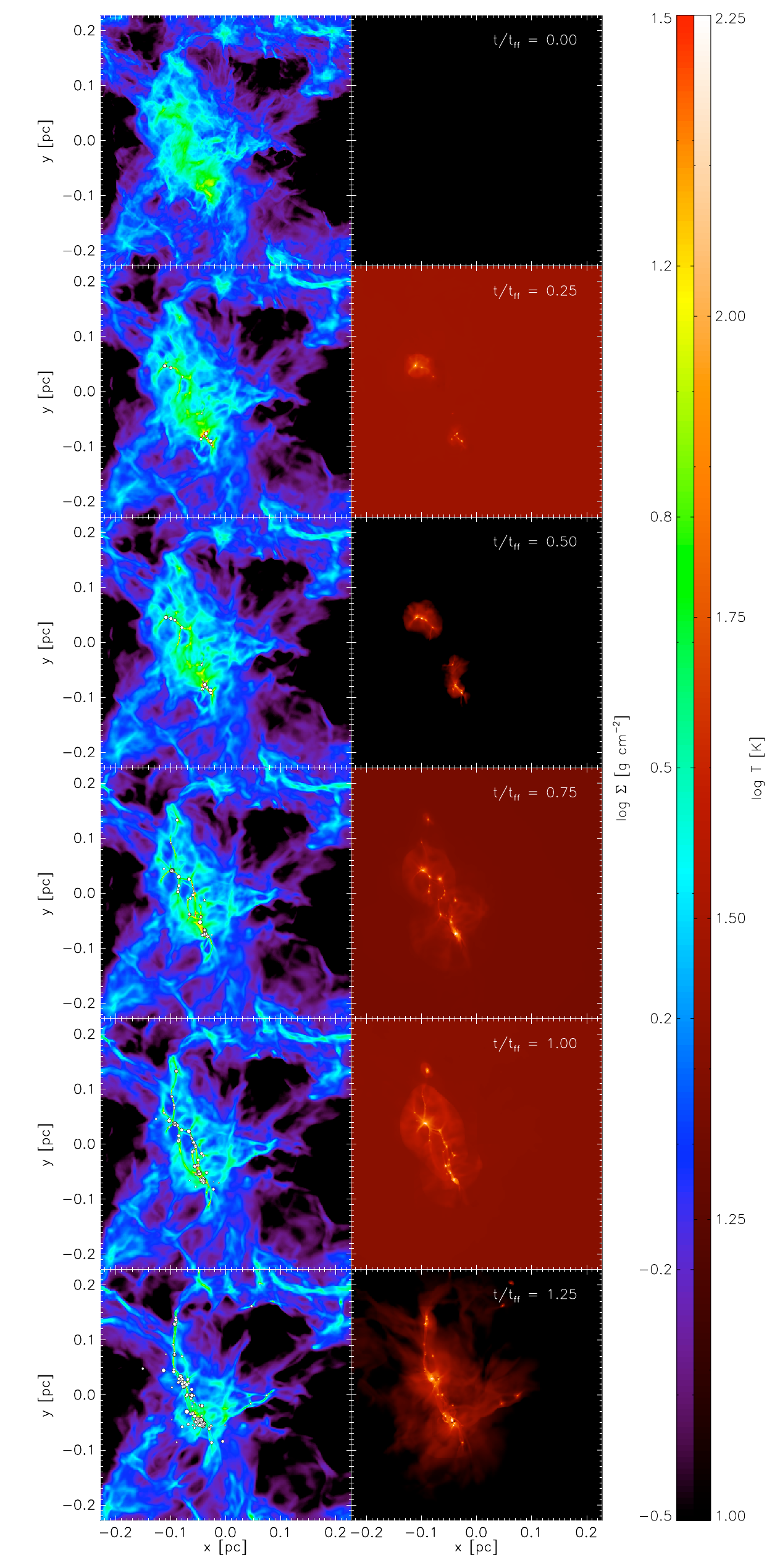}
\caption{
\label{fig:snapshots_tunw}
Same as Figure \ref{fig:snapshots_smnw}, but for run TuNW. Note that the color scales are the same, but the size of the region shown is slightly different.\\
}
\end{figure}

\begin{figure}
\plotone{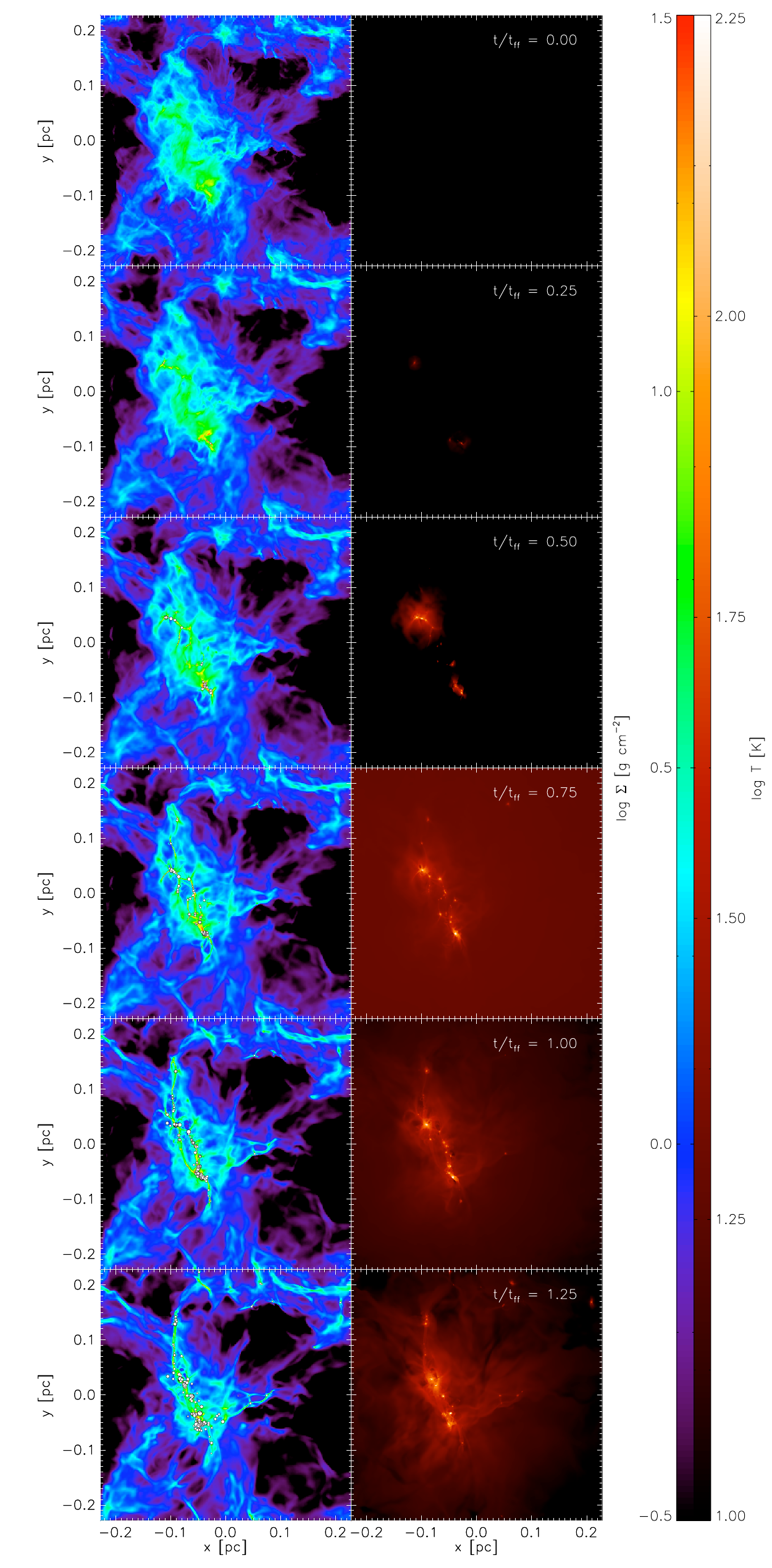}
\caption{
\label{fig:snapshots_tuw}
Same as Figure \ref{fig:snapshots_tunw}, but for run TuW.
}
\end{figure}

In Figures \ref{fig:snapshots_smnw}, \ref{fig:snapshots_tunw}, and \ref{fig:snapshots_tuw} we show the large-scale column density and density-weighted temperature distributions in for runs SmNW, TuNW, and TuW, respectively.  In all three we observe the same general trend: the turbulence creates an overdense region, which then begins to collapse and form stars. The collapsing structures are filamentary, and the stars are born along the filaments, and particularly at the nodes where the filaments intersect. The temperature is initially small, but as stars form hot spots around individual stars appear, and these gradually spread over time.\footnote{In runs TuNW and TuW there are sometimes brief increases in the overall background temperature level visible in some snapshots, but these are short-lived and small, generally keeping the temperature $<20$ K. These flashes are associated with brief increases in the accretion luminosity that are large enough to heat the entire simulation volume above 10 K for short periods.}

There are a few interesting points to take from these plots. One involves the morphology of the heated regions. In the turbulent runs, even at late times the temperature distribution looks more like a series of islands of heated gas surrounded by a large medium that is either at or quite near to the background temperature. In effect, one can discern something like individual protostellar cores that are heated by the star or star system embedded within them. In contrast, by the end of run SmNW there is simply a single, concentrated region of heating, and one cannot discern individual cores any more. As we show below, this difference proves to be important in determining the evolution of the IMF.

Another interesting point is that the overall morphology is surprisingly similar in runs TuW and TuNW, despite the change in whether we include protostellar winds or not. Partly this is a function of the fact that wind-blown bubbles are fairly low column-density structures, and that we are looking at static slices. In an animation of the column density field, one readily discern outflows driving shells of gas orthogonal to the filaments. However, this clearly has a relatively small effect on the large-scale morphology.

\subsection{Star Formation Rate and History}

\begin{figure}
\epsscale{1.1}
\plotone{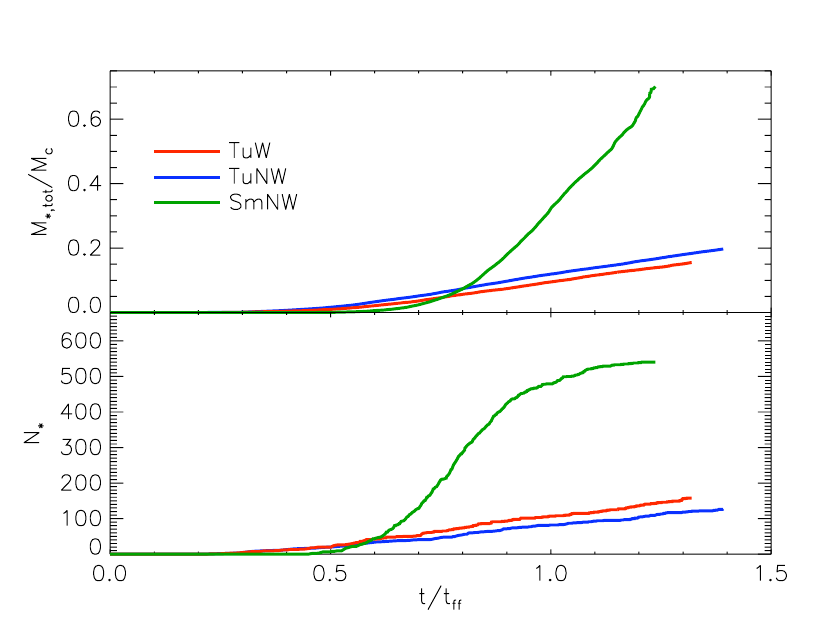}
\caption{
\label{fig:starhist1}
Total mass in stars (top) and total number of stars (bottom) as a function of time in runs SmNW, TuNW, and TuW.
}
\end{figure}

In Figure \ref{fig:starhist1} we show the star formation history of each of our simulations. The most immediate and striking thing about the Figure is the difference in star formation histories between the smooth and turbulent runs. Run SmNW starts off its star formation more slowly than TuNW or TuW, which is not surprising since its initial density structure is smooth, and possesses no high density peaks that collapse quickly. However, star formation in that run accelerates dramatically as the time approaches $\tff$. 
\red{Late in the simulation, the star formation rate approaches $\sim 1-2 M_c/\tff$.}
In contrast, in runs TuNW and TuW the star formation rate is roughly constant and fairly low. After a time $\tff$, only about 10\% of the mass has been turned into stars. There is no obvious acceleration with time. Note that, although part of the difference in star formation rates comes from the difference in free-fall times between the turbulent and smooth runs, even if we were to measure in seconds rather than free-fall times run SmNW would have a much larger star formation rate than TuNW or TuW. We summarize the dimensional and dimensionless star formation rates in the simulations in Table \ref{tab:runoutcome}. Observationally, the dimensionless star formation rate \citep{krumholz05c}
\begin{equation}
\epsilon_{\rm ff} \equiv \frac{\dot{M}_*}{(1/2) M_c/\tff},
\end{equation}
where the factor of $1/2$ arises because half the cloud mass is above the density $\langle \rho \rangle_M$ used to define $\tff$,
is $\sim 1\%$, with roughly half a dex scatter, over a very wide range of densities and galactic environments \citep{krumholz07e, evans09a, krumholz12a}.\footnote{There is some subtlety in the observational comparison here, because real observations usually have an upper limit on the density to which they are sensitive, for example because the tracer being used depletes or becomes very optically thick at high density. Since we include all the mass above $\langle \rho\rangle_M$ in our computation of $\epsilon_{\rm ff}$, we are not capturing this effect. However, the change in mass it would induce is small, because both real star-forming clouds and our simulated turbulent clouds have density probability distribution functions that are sharply declining at densities above the peak. Thus the amount of mass missed due to the density upper limit in the observations is likely to be very small.} Comparing to the table, we see that $\epsilon_{\rm ff}$ in runs TuNW and TuW is still roughly an order of magnitude too high compared to observations, but it is roughly an order of magnitude lower than in run SmNW. We discuss the origin of the remaining discrepancy between TuW and the observations further in Section \ref{sec:implications}.

\begin{deluxetable}{cccccc}
\tablecaption{Simulation Outcomes\label{tab:runoutcome}}
\tablehead{
\colhead{Name} &
\colhead{$t_{\rm fin}/\tff$}  &
\colhead{$M_{*,\rm fin}/M_c$} &
\colhead{$N_*$} & 
\colhead{$10^3 \dot{M}_*$} &
\colhead{$\epsilon_{\rm ff}$}
\\
\colhead{} & 
\colhead{} &
\colhead{} &
\colhead{} &
\colhead{ ($\msun$ yr$^{-1}$)} & 
\colhead{}
}
\startdata
SmNW & 1.25 & 0.70 & 540 & 16 & 1.78 \\
TuNW & 1.39 & 0.20 & 127 & 7.4 & 0.33 \\
TuW & 1.32 & 0.15 & 158 & 6.2 & 0.28 \\
\enddata
\tablecomments{Col.\ 2: time at which run was stopped. Col.\ 3: total stellar mass at the end of the run. Col.\ 4: number of stars present at the end of the run. Col.\ 5: time-averaged star formation rate in the run, measuring from formation of the first star to the end of the run. Col.\ 6: dimensionless star formation rate $\epsilon_{\rm ff} \equiv \dot{M}_* / [(1/2)M_c/\tff]$.
}
\end{deluxetable}

The difference in star formation rate (SFR) between the runs may be understood readily if we consider what happens to the turbulence, which, in these runs with no magnetic fields, is the main mechanism for regulating the SFR. In run SmNW, the initial turbulence present in the gas decays, and after one crossing time, which is $\sim \tff$, this decay gives rise to a global collapse and an accelerating star formation rate. In contrast, for runs TuNW and TuW, the box crossing time, and thus the turbulent decay time, is significantly longer than the free-fall time at the mass-weighted mean density. It is comparable to the free-fall time at the volume-weighted mean density, which is much longer. The difference in star formation history between the runs makes a critical point: it matters for their star formation rates that star-forming dense clumps like the one out of which the ONC formed are not isolated objects. They are instead the inner parts of larger turbulent structures, and the energy from those larger scales is able to cascade down to smaller scales and maintain the turbulence for longer than the dynamical times of the small clumps. The turbulent decay timescale in a proto-ONC gas clump is the crossing time of its parent molecular cloud, not the crossing time of the small clump. This point has previously been made by \citet{falceta-goncalves11a} in the context of non-self-gravitating turbulence, and our work strongly confirms their conclusion and extends it to the self-gravitating case.

In contrast, the differences between the two turbulent runs are relatively small. The star formation rate measured by mass (as opposed to number of stars) is $\sim 20\%$ lower in TuW than in TuNW. Since our model for protostellar outflows prescribes that 27\% of the mass that reaches a star particle (and thus the inner wind launching region) be ejected in an outflow, this reduction in the star formation rate is surprisingly small. This implies that there must be very little entrainment of additional material by the outflows, and even that some of the material that is entrained by outflows must be rapidly stopped and recycled back into the star-forming region. Visual inspection of the morphology of the outflows and accretion flows confirms that this is in fact the case: outflow shocks visible in the animations are generally traveling at right angles to the filaments feeding the stars. This is not an accident. Each star launches its bipolar outflow along the axis specified by its angular momentum vector. If stars are being fed primarily by filaments lying in a plane, as is the case in all our runs, then most of their angular momentum vectors tend to be perpendicular to that plane, producing relatively little entrainment. The minority of outflows that do end up aligning with the filaments possess too little momentum to significantly hinder the accretion flow, and the matter they do eject is stopped by the greater ram pressure of the infalling gas. As a result, it is re-accreted fairly rapidly. Whether this behavior is actually realistic is a separate question, one to which we return in Section \ref{sec:discussion}.

\subsection{The IMF}

\begin{figure}
\epsscale{1.1}
\plotone{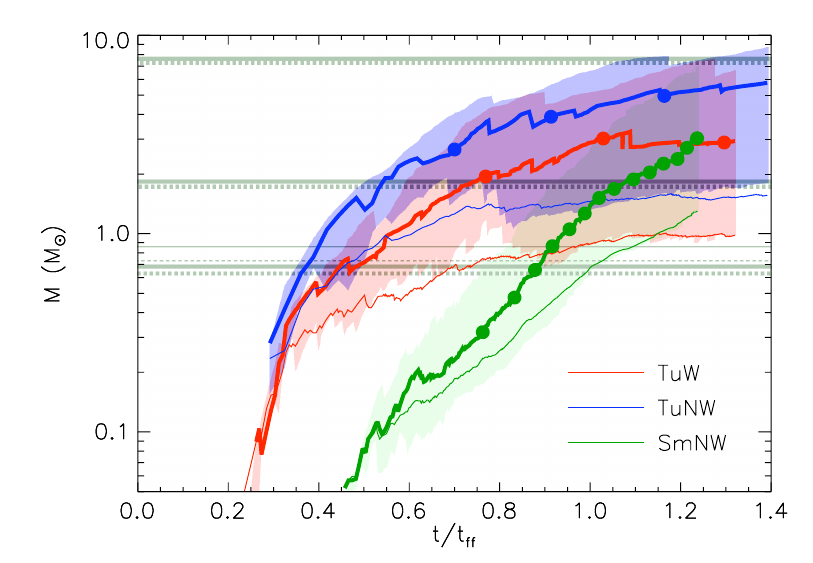}
\caption{
\label{fig:mperc}
Evolution of the IMF over time in the three simulations. Thick lines indicate the 50\% percentile mass $M_{50}$ (see main text for formal definition), while the shaded regions indicate the range between the 25th and 75th percentile masses $M_{25}$ and $M_{75}$. Thin lines indicate the mean mass $\overline{M}$. Colors indicate the run, as described in the legend. Circles long the thick lines indicate the points at which the stellar mass reaches 50 $\msun$, 100 $\msun$, 150 $\msun$, etc. For comparison, thick gray unbroken horizontal lines show $M_{25}$, $M_{50}$, and $M_{75}$ for a fully sampled \citet{da-rio12a} IMF (Equation \ref{eq:darioimf}), and the thin gray unbroken horizontal line shows $\overline{M}$ for this IMF. Dashed lines show the equivalent quantities for a \citet{chabrier05a} IMF.\\
}
\end{figure}

Another interesting feature in Figure \ref{fig:starhist1} is that the mass in stars in run TuW is $\sim 20\%$ smaller than in TuNW at equal times, but that the number of stars is $\sim 20\%$ larger in TuW. This indicates an important shift in the stellar IMF between the runs. Although it is less obvious visually from Figure \ref{fig:starhist1}, there are also very important differences in the IMF between run SmNW and the other two runs. We now examine these.

For the purposes of quantitative comparison between different simulations, and between simulations and observations, it is helpful to examine percentiles in the cumulative mass distribution function for stars produced in the runs. We define the $n$th percentile mass $M_n$ implicitly via the equation
\begin{equation}
\sum_{m_* < M_n} m_* = \frac{n}{100} \sum m_*,
\end{equation}
where $m_*$ is the mass of each individual star, the first sum runs over stars with masses $m_* < M_n$, and the second sum rus over all stars. Thus, for example, $M_{50}$ is defined by the condition that sum of the masses of all stars smaller than $M_{50}$ constitutes 50\% of the total stellar mass. We also examine the mean stellar mass, defined by
\begin{equation}
\overline{M} = \frac{\sum m_*}{N_*}
\end{equation}
where $N_*$ is the total number of stars. We can measure each of these quantities directly from our simulations at every time. We can also compare the simulation IMFs to observed ones. We select two observational IMFs for comparison. In ONC, \citet{da-rio12a} find for low mass stars an IMF well-fit by a lognormal function with a width of $\sigma=0.44$ in $\log m_*$, centered on $\log m_{*,c} = -0.45$ (measured in $\msun$; their Table 3). The highest mass bin in \citeauthor{da-rio12a}'s sample is $\sim 2$ $\msun$, so to extend this to higher masses we adopt a \citet{chabrier03a} functional form in which the lognormal at low mass has a powerlaw tail of slope $-1.35$ at high mass. Thus the observed ONC IMF to which we compare is
\begin{equation}
\label{eq:darioimf}
\frac{dN}{d\log m_*} \propto
\left\{
\begin{array}{ll}
e^{-(\log m_*-\log m_{*,c})^2/2\sigma^2}, & m_* < \msun \\
e^{-\log m_{*,c}^2/2\sigma^2} m_*^{-1.35} & m_* \ge \msun,
\end{array}
\right.
\end{equation}
where all masses are in Solar units, over a range from $0.05 - 150$ $\msun$. For this IMF, $M_{25} = 0.69$ $\msun$, $M_{50} = 1.8$ $\msun$, $M_{75} = 7.6$ $\msun$, and $\overline{M} = 0.86$ $\msun$. The second comparison IMF is the system IMF of \citet{chabrier03a, chabrier05a} for the galactic field, which also seems to fit other star clusters reasonably well \citep{parravano11a}. We use the system rather than the single star IMF because we do not resolve tight binaries. This IMF has the same functional form as Equation (\ref{eq:darioimf}), but with $\log m_{*,c} = -0.60$ and $\sigma= 0.55$. The corresponding percentile and mean values are $M_{25} = 0.63$ $\msun$, $M_{50} = 1.7$ $\msun$, $M_{75} = 7.2$ $\msun$, and $\overline{M} = 0.73$ $\msun$. The \citeauthor{chabrier05a} and \citeauthor{da-rio12a} IMFs differ significantly in the number of brown dwarfs and very low mass stars they predict, but converge at masses above a few tenths of $\msun$. For a discussion of possible origins of the discrepancy between the two IMFs, we refer readers to \citet{da-rio12a}.

In Figure \ref{fig:mperc} we plot the time evolution of $M_{25}$, $M_{50}$, $M_{75}$, and $\overline{M}$ in each of our simulations, and for the observed \citet{da-rio12a} and \citet{chabrier05a} IMFs. The figure immediately reveals some interesting results. First, we see that in run TuW the IMF is in remarkably good agreement with the observed IMFs. At the end of the simulations, the mean mass $\overline{M}$ agrees with the observed \citeauthor{da-rio12a} value to better than 20\%, and the 50th percentile mass $M_{50}$ to less than a factor of 2; we show below that this level of disagreement is consistent with coming simply from statistical sampling variance. Moreover, and perhaps more importantly, the agreement is good at almost all times when there is a significant mass of stars present, because the IMF in run TuW is very stable over time. From $\sim 0.7 \tff$, when the total stellar mass reaches 50 $\msun$, to $\sim 1.3 \tff$, when it reaches $150$ $\msun$, we find that $M_{50}$, $M_{25}$, and $\overline{M}$ stay constant to within $\sim 50\%$; $M_{75}$ changes slightly more, almost certainly as a result of under-sampling the high end of the IMF when there are relatively few stars. The change becomes even smaller at later times. From the time when 10\% of the mass is in stars ($t\sim  1.0\tff$) to when 15\% is in stars, $M_{50}$ changes by less than 5\% and $\overline{M}$ by less than 10\%.

In contrast, for run TuNW the mean mass is relatively stable, but $M_{50}$ rises systematically with time, increasing by a factor of 2.2 as the stellar mass grows from $50$ to $200$ $\msun$, corresponding to times $t/\tff \sim 0.7$ to $1.4$. This reflects more rapid growth of the more massive stars in the run where winds do not suppress accretion. Moreover, in this run the rate at which new stars form is lower than in run TuW. The agreement with observations in this case is clearly weaker; the run produces an IMF that is too top-heavy.

The changes with time in run TuNW, however, are small compared to those that occur in run SmNW. There $M_{50}$ and $\overline{M}$ increase by nearly an order of magnitude in a time less than $0.5 \tff$. Each increase in stellar mass of $50$ $\msun$ is accompanied by a factor of $\sim 2$ gain in $M_{50}$. This pattern of growth occurs due to the ``overheating" problem discussed in Paper I: in run SmNW, star formation is much too rapid and too concentrated, and this produces a rapidly rising accretion luminosity that heats the gas mass to the point where the Bonnor-Ebert mass is too large for stars for new small stars to form. Accretion continues, but it is entirely captured by the existing stellar population, leading to an IMF whose mean and median mass rise with time. Moreover, since all the stars are growing in lockstep the mass distribution in this run is too narrow as well.

\begin{figure}
\epsscale{1.1}
\plotone{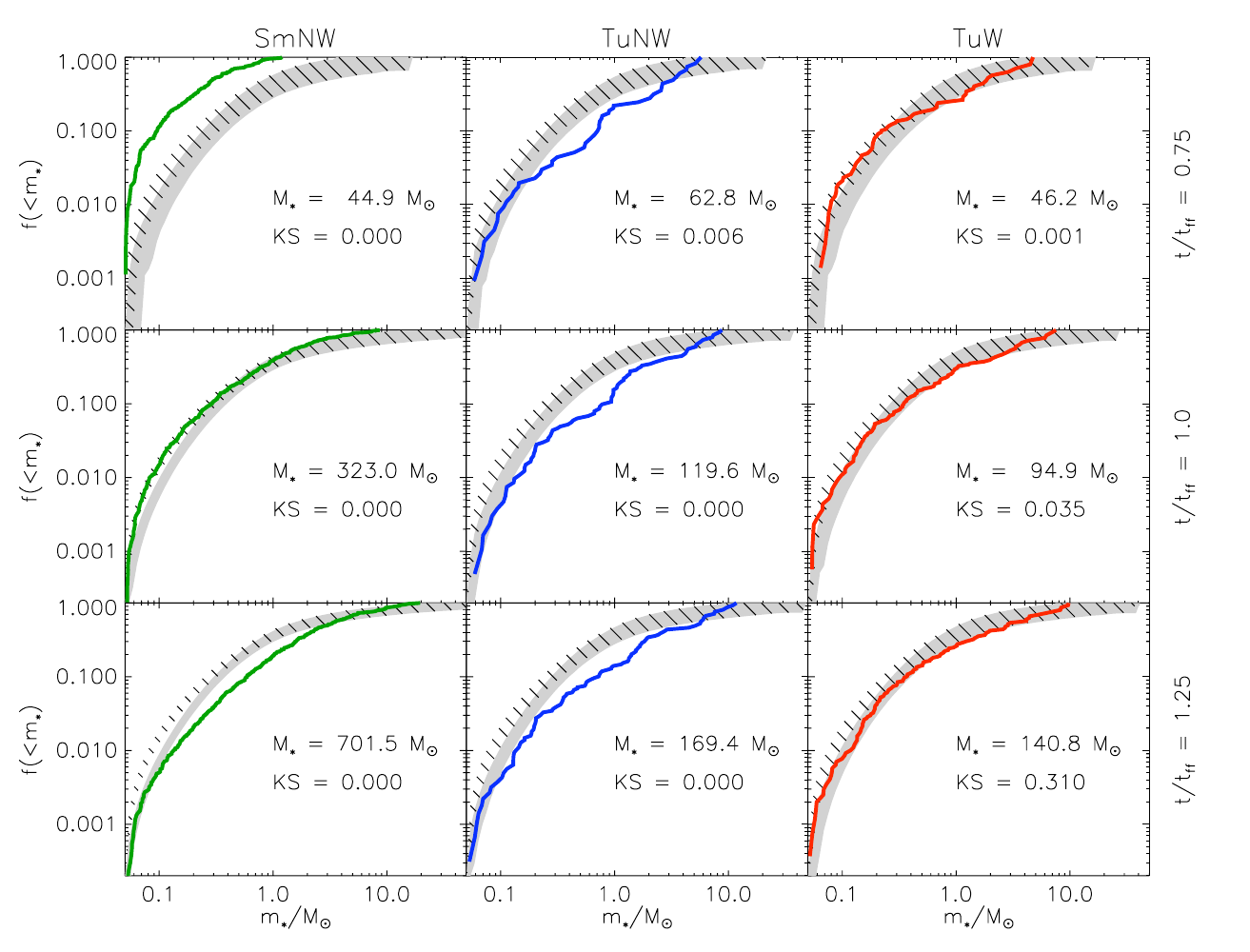}\\
\plotone{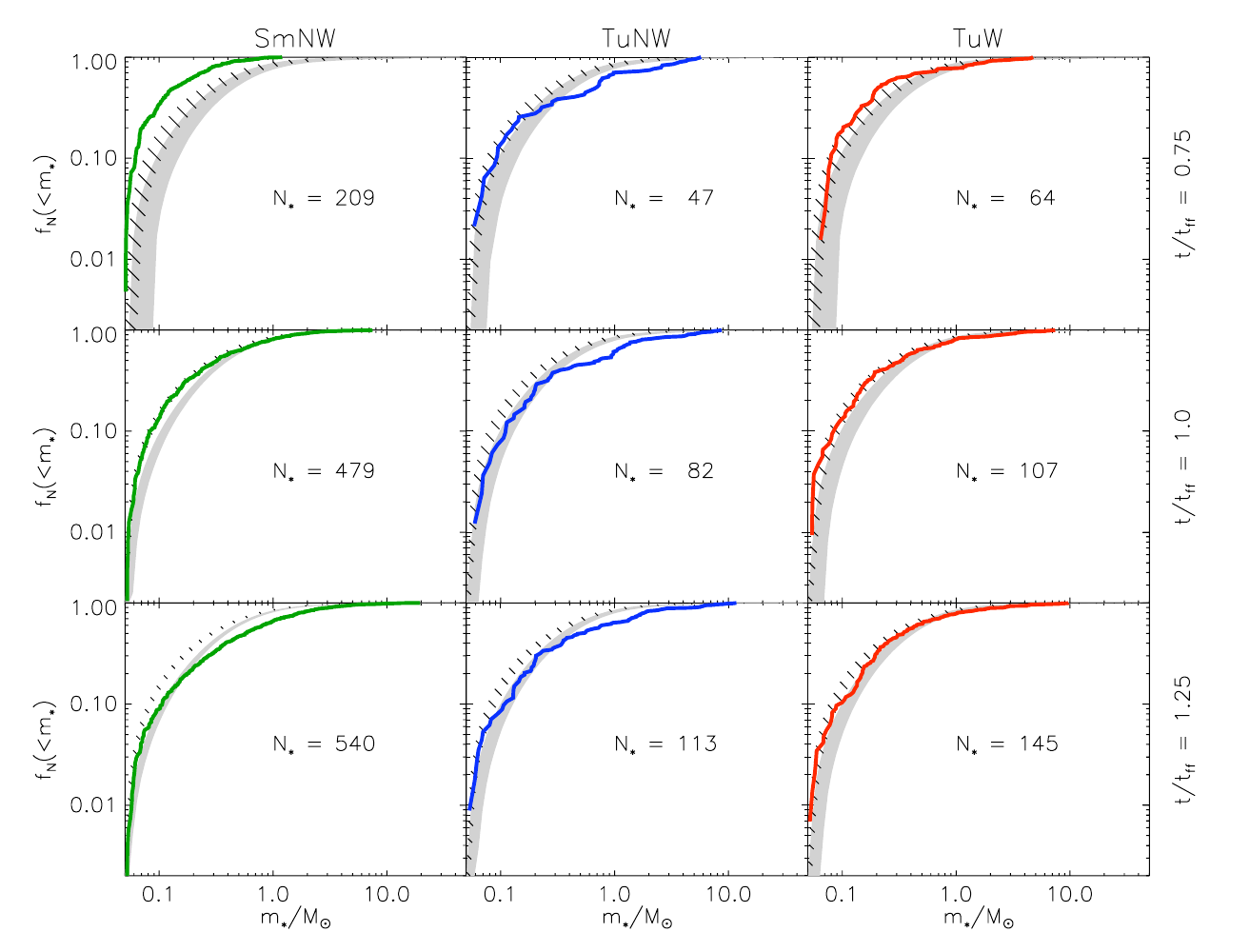}
\caption{
\label{fig:imfplot1}
Cumulative mass functions in the simulations, compared to observations. \red{The top set of panels shows the cumulative distribution by mass, and the bottom shows the distribution by number. Within each set of panels,} columns show the results from runs SmNW, TuNW, and TuW, as indicated. Rows correspond to times $t/\tff = 0.75, 1.0,$ and 1.25, as indicated. In each panel, the colored line indicates the fraction of stellar mass $f_M(<m_*)$ \red{or the fraction of the number of stars $f_N(<m_*)$} in stars with mass less than $m_*$ in the simulation, and the gray band indicates the range from the 10th to 90th percentile resulting from drawing a large number of clusters from the \citet{da-rio12a} IMF. The hatched band is the 10th to 90th percentile range for a \citet{chabrier05a} IMF. For details on how this drawing is done, see the Appendix to Paper I. The label in each panel indicates the total mass \red{or number} of stars at that time in that simulation.\\
}
\end{figure}

\begin{figure}
\epsscale{1.1}
\plotone{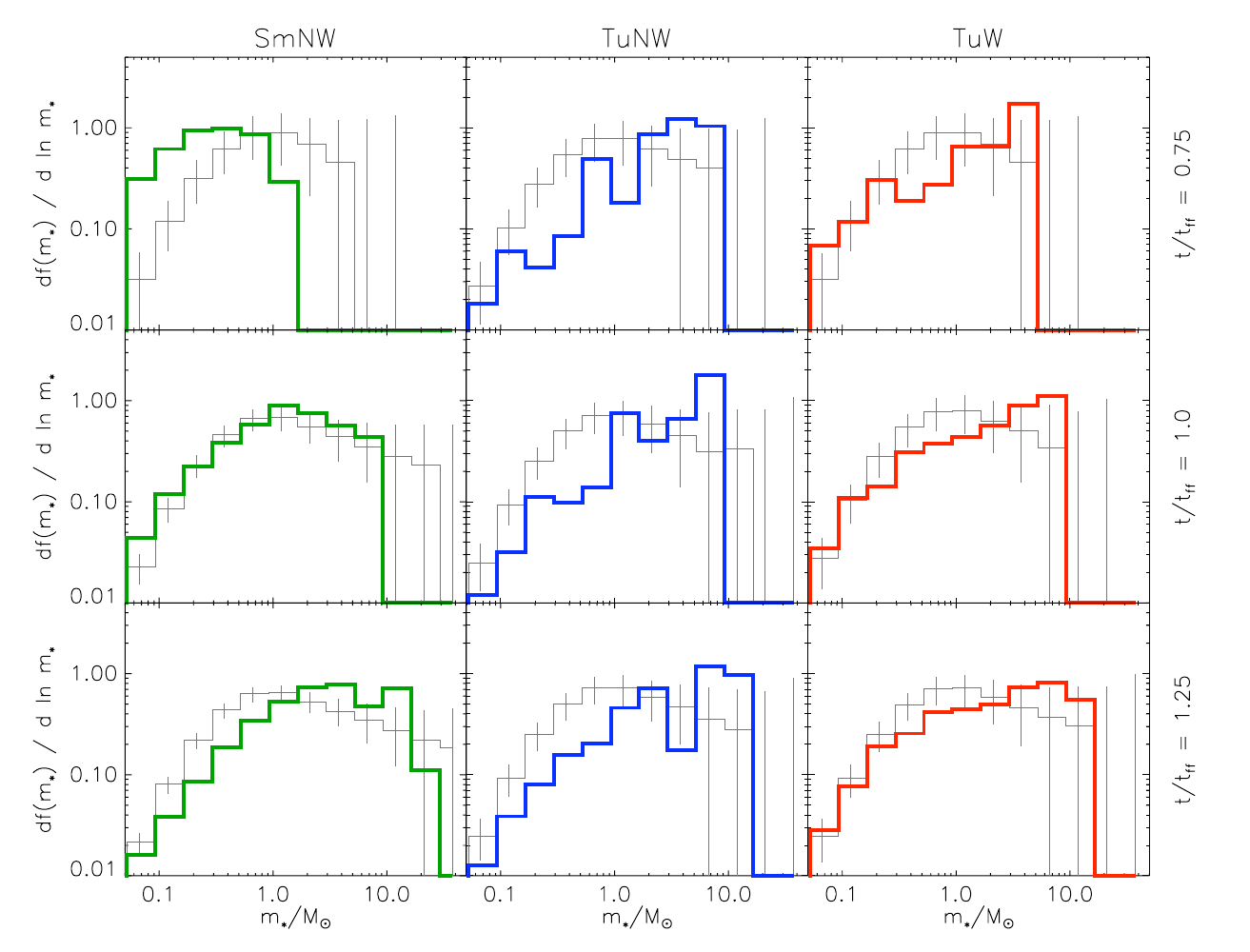}\\
\plotone{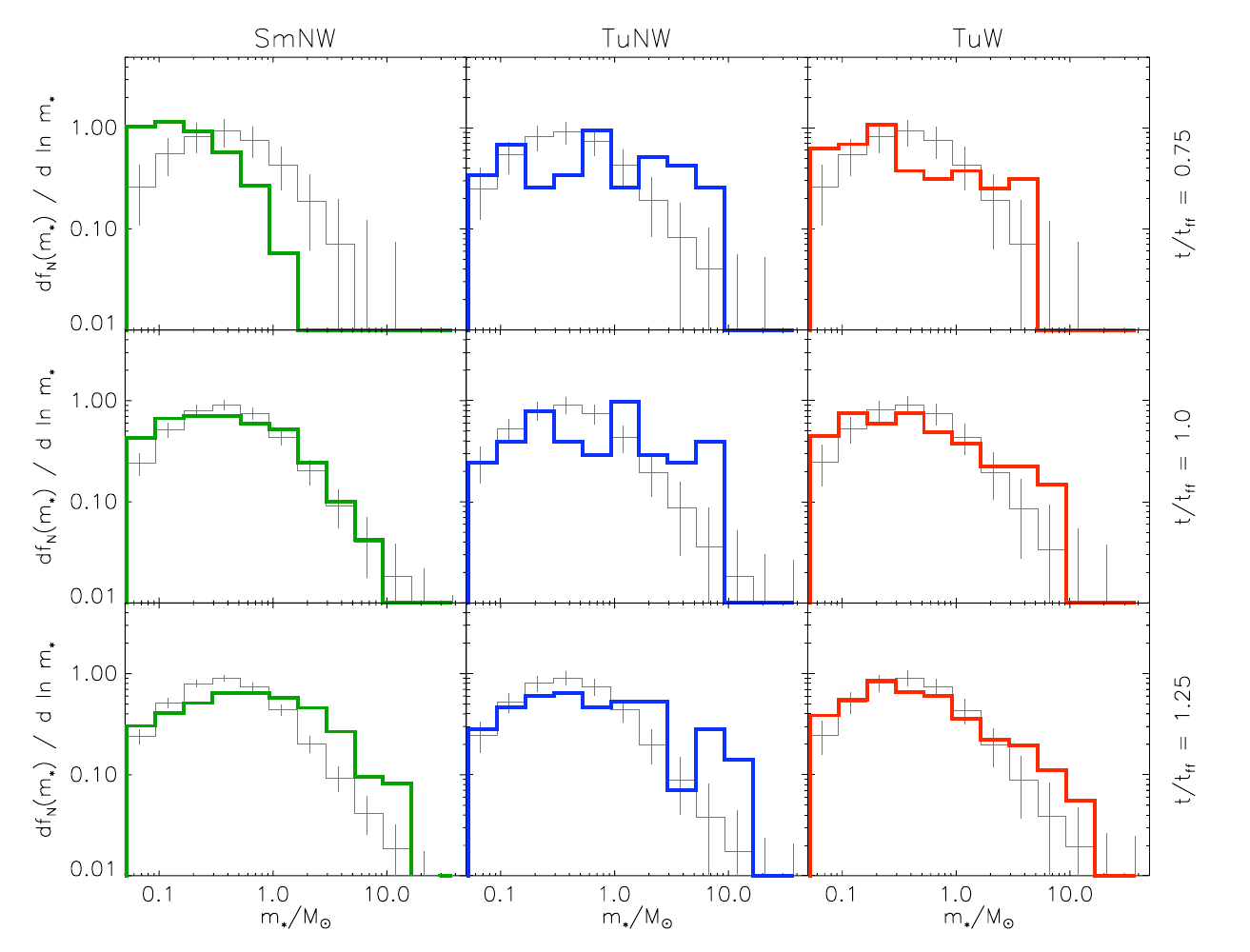}
\caption{
\label{fig:imfplot2}
Same as Figure \ref{fig:imfplot1}, but showing differential rather than cumulative mass distributions. The histogram value in each bin shows the total fraction of all stellar mass \red{(for the top panels) or the total fraction of the number of stars (for the bottom panels)} falling within that bin. Thick colored lines indicate the simulation result, and gray lines indicate the results of drawing an equal mass stellar population from the \citet{da-rio12a} IMF. For the gray histogram, the histogram values give the median result, and the vertical lines indicate the range from the 10th to the 90th percentile. We omit the \citet{chabrier05a} IMF here to reduce clutter.\\
}
\end{figure}

\begin{figure}
\epsscale{1.1}
\plotone{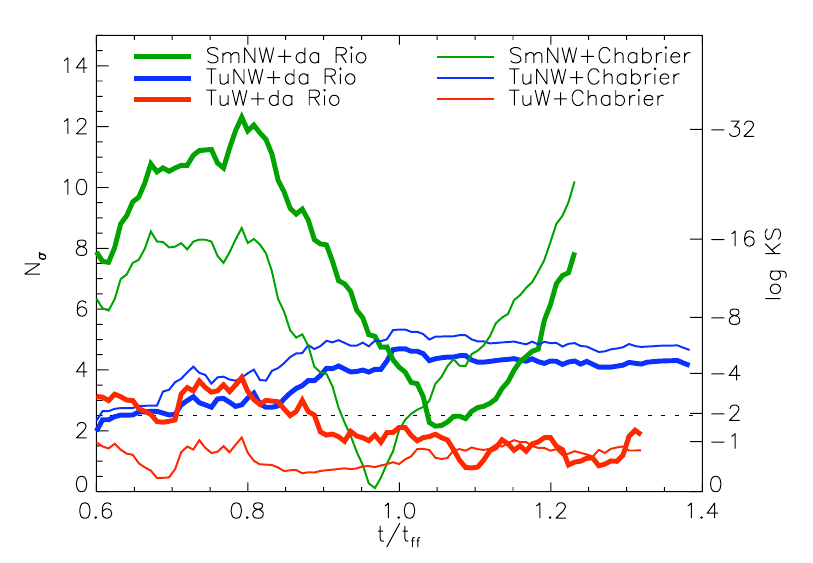}
\caption{
\label{fig:imfplot_ks}
\red{Level of statistical agreement between the simulation and observed IMFs as a function of time. At each time, the quantity plotted is the result of a KS test comparing the three simulations (green for SmNW, blue of for TuNW, and red for TuW) to the \citet[thick lines]{da-rio12a} and \citet[thin lines]{chabrier05a} IMFs. The right axis shows the $P$-value returned by the KS test, where $1-P$ is the confidence level at which we can rule out the null hypothesis that the simulation and observed IMFs are drawn from the same underlying distribution. The left axis shows the equivalent confidence level measured in number of standard deviations, which is related by $N_\sigma = \sqrt{2}\, \mbox{erfc}^{-1}(P)$, with $\mbox{erfc}$ the complementary error function. The dashed horizontal black line indicates a confidence level of $2.5\sigma$.\\
}
}
\end{figure}

\red{We can also make this comparison more quantitatively.} In Figures \ref{fig:imfplot1} and \ref{fig:imfplot2} we show the cumulative and differential mass distributions produced in our simulations at various times, and compare to observed IMFs. \red{At each time in the simulations, we can quantitatively described the level of consistency or inconsistency between the simulated and observed IMFs using a Kolmogorov-Smirnov (KS) test. We plot the result in Figure \ref{fig:imfplot_ks}. Examining the figures, we see that run SmNW is strongly inconsistent with both observed IMFs at most times. At early times the IMF is too bottom-heavy, but as time increases the IMF peak shifts to higher masses. Around $t/\tff = 1$ run SmNW is fully consistent with the Chabrier IMF, and marginally consistent with the da Rio one, but at later times the IMF peak continues to shift to higher values and becomes inconsistent with both once more. This is the overheating problem described in Paper I. For run TuNW the IMF peak does not shift systematically with time, and so there is no overheating problem. However, the absolute value of the mean mass is systematically too high, as shown in Figure \ref{fig:mperc}. As a result, the overall level of agreement between the simulation and the observed IMFs is poor. On the other hand, run TuW is generally statistically consistent with both the da Rio and Chabrier IMFs at most times.}

\red{It is important to add some caveats to this result. First, the KS statistic does not account for the observational and systematic uncertainties in the observed IMFs. Were these uncertainties to be included, it is entirely possible that run TuNW would be consistent within them, and perhaps even that run SmNW would be, at least for a longer period of time. Second, the KS test itself is an imperfect tool. It is most sensitive to differences in distributions near the 50th percentile, and less sensitive to differences on the tail of the distribution. Thus, for example, Figure \ref{fig:imfplot2} shows that run SmNW has a slight excess of stars in the $\sim 3-10$ $\msun$ range that is clearly visible in a differential mass function on a logarithmic axis. The KS test does not regard this excess as statistically significant, but it is conceivable that a more sensitive statistical test might. Indeed, one can get a sense of the level of statistical power that the KS test provides when applied to our simulations from the fact that, formally, our simulations are consistent with both the da Rio and Chabrier IMFs. This is partly because we are not performing a comparison in the mass range $0.01 - 0.05$ $\msun$ where the two distributions are most different, but it is also partly because, with only 158 stars in run TuW, there is significant sampling noise.
}

%

\subsection{Gas Thermodynamics}

\begin{figure*}
\epsscale{1.1}
\plotone{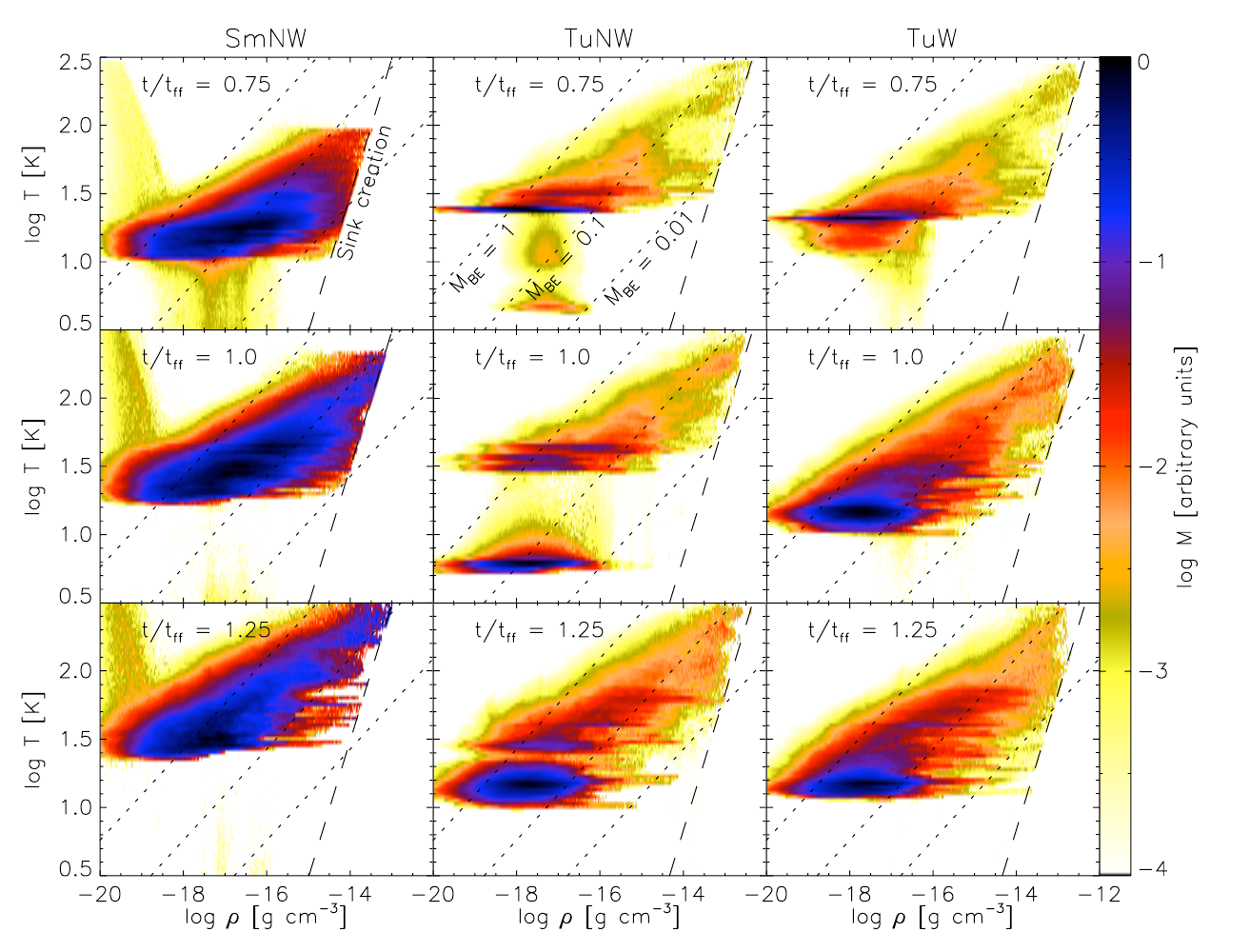}
\caption{\label{fig:phaseplot}
Phase diagrams of the three runs at different times. The three columns correspond to runs SmNW, TuNW, and TuW, as indicated. The three rows correspond to times $t/\tff=0.75$, $1.0$, and 1.25. In each panel, the color indicates the gas mass in a given bin of density and temperature; bins are 0.025 dex wide in both $\rho$ and $T$. The color scale is normalized so that the bin containing the largest amount of mass is 1.0.  The long-dashed line indicates the locus in density and temperature at which the code inserts sink particles. The short-dashed lines indicate the locus in density and temperature where the Bonnor-Ebert mass is $0.01$ $\msun$, $0.1$ $\msun$, and $1$ $\msun$ as indicated. Note that gas in the winds is run TuW is heated to $\sim 10^4$ K, well above the temperature range shown here, but there is relatively little mass at these temperatures.\\
}
\end{figure*}

The fragmentation of the gas is driven by its thermodynamics, and we can gain insight into the differences in outcome between the runs by examining the temperature structure of the gas. In Figure \ref{fig:phaseplot} we show phase diagrams of the three runs at three different times. Not surprisingly, each of the runs is quite different. First examining run SmNW, we note that, at time $t/\tff = 0.75$, the bulk of the gas in run SmNW is cooler than in the other two runs. This reflects the fact that the total stellar mass in run SmNW is comparable to that in runs TuNW and TuW at this point. Since the free-fall time is longer in run SmNW, this corresponds to a lower total accretion rate and thus a lower accretion luminosity. However, as star formation in run SmNW accelerates, the accretion luminosity rises and the gas heats, while the gas in the other two runs stays relatively cool. Quantitatively, at the final times shown in the bottom row of Figure \ref{fig:phaseplot}, 42\% of all the gas is at temperatures above 50 K in run SmNW (excluding the ambient medium); the equivalent Figures in both runs TuNW and TuW are 7\%. It is important to note that this difference is driven by accretion luminosity and not by the intrinsic luminosity of massive stars. If we instead examine run SmNW at time $t/\tff = 1.0$, the most massive star present is 8.8 $\msun$, smaller than the most massive stars present at time $t/\tff = 1.25$ in runs TuNW (13.3 $\msun$) and TuW (9.9 $\msun$). Nonetheless, we still find that 23\% of the mass is at temperatures above 50 K, and 64\% is above 30 K, i.e.\ there is more hot gas in run SmNW even when the individual stars are less massive.

The rapid heating in run SmNW gives rise to the overheating problem identified in Paper I -- bulk heating of all the gas makes it impossible for small stars to form, thus shifting the IMF systematically to higher mass as time goes on. Runs TuNW and TuW clearly do not suffer from this problem. Even at late times, the great majority of their gas is at temperatures of no more than $10-15$ K, and there is very little material at temperatures of more than 50 K. Although there clearly is gas being warmed by stars in these runs, there remain pockets of cold, gas at densities $> 10^{-15}$ g cm$^{-3}$ and temperatures $<15$ K that is capable of producing new stars with masses $\sim 0.01$ $\msun$. These are visible in Figure \ref{fig:phaseplot}, where the phase diagram reveals the presence of material for which the Bonnor-Ebert mass,
\begin{equation}
M_{\rm BE} = 1.18 \frac{c_s^3}{\sqrt{G^3 \rho}}
\end{equation}
is below $0.01$ $\msun$. In contrast, at late times in run SmNW there no material for which $M_{\rm BE}$ is this small. To be quantitative, at the times shown in the final panel of Figure \ref{fig:phaseplot}, run SmNW contains only $1.8\times 10^{-3}$ $\msun$ of material in the density and temperature region where $M_{\rm BE} < 0.01$ $\msun$, i.e.\ too little mass to actually create a star. The corresponding figures for runs TuNW and TuW are $0.49$ and $1.0$ $\msun$, respectively, 
\red{making it possible for new brown dwarfs to form.}
It is interesting to note that the amount of cold, high-density gas is generally greater in run TuW than in run TuNW. This is likely an effect of the reduced accretion rate and changed IMF in run TuW compared to TuNW, both of which serve to generally lower the accretion luminosity and thus the heating rate. \red{The spatial distribution of the star formation may also play a role: in run SmNW, because there is no pre-existing density structure at the start of the simulation and because the turbulence decays rapidly, all the stars and gas become concentrated in a single dominant cluster, where stellar heating is very intense. In runs TuNW and TuW, the combination of a pre-existing density structure present in the initial conditions and the non-decay of turbulence throughout the simulation serves to break star formation up into several subclusters, within each of which stellar heating is less intense.}

\subsection{Massive Cores and Massive Stars}

\begin{figure*}
\epsscale{1.1}
\plotone{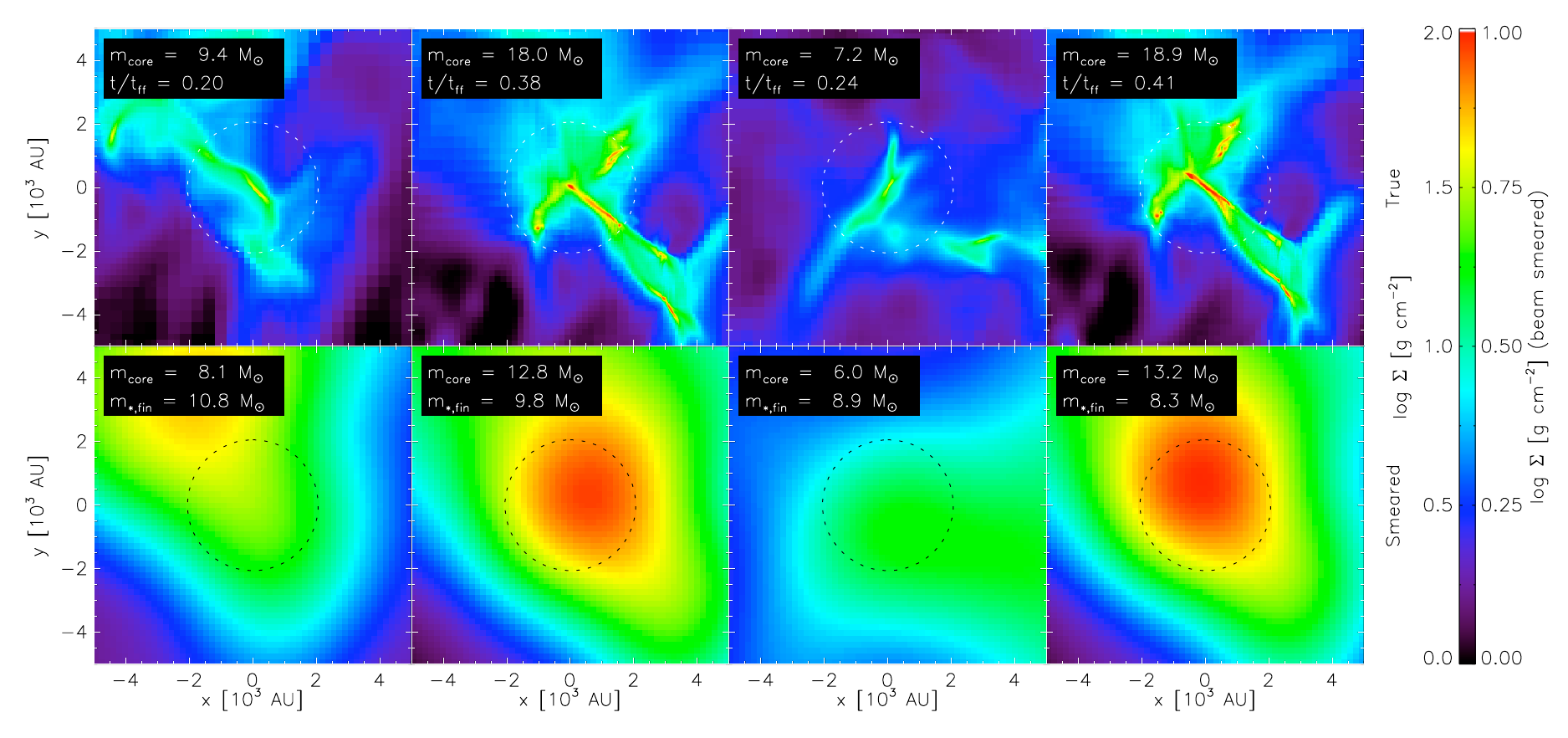}
\caption{\label{fig:snapshots_hmcores}
Images of the initial cores that produced the four most massive stars in simulation TuW. In each column, the upper image shows the column density distribution, centered on the $\sim 0.05$ $\msun$ protostars that will grow to be massive stars. The lower image shows the same column density distribution, smeared with a $1700$ AU Gaussian beam. In the upper panels we indicate the mass of the core (defined as the projected mass within a radius of $0.01$ pc, as indicated by the dashed circles) and the time at which the snapshot is taken. In the lower panels we indicate the core mass that would be inferred from the beam-smeared image and the final mass of the resulting star. Note that the second and fourth columns are nearly identical because two of the final massive stars both form in the same core.
}
\end{figure*}

\begin{figure*}
\epsscale{1.1}
\plotone{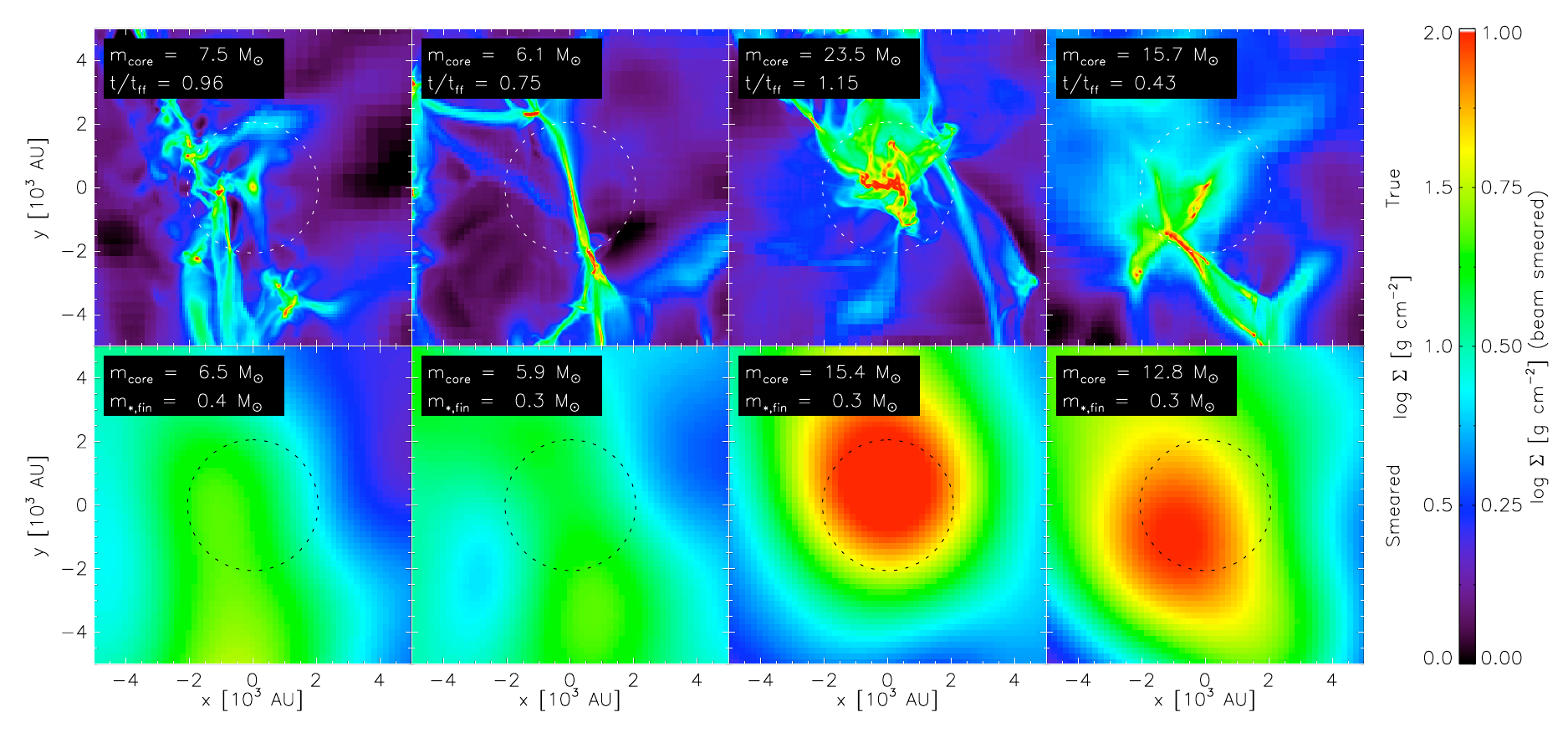}
\caption{\label{fig:snapshots_lmcores}
Same as Figure \ref{fig:snapshots_hmcores}, but for the four stars closest to the median of the final mass distribution.\\
}
\end{figure*}

\begin{figure}
\epsscale{1.0}
\plotone{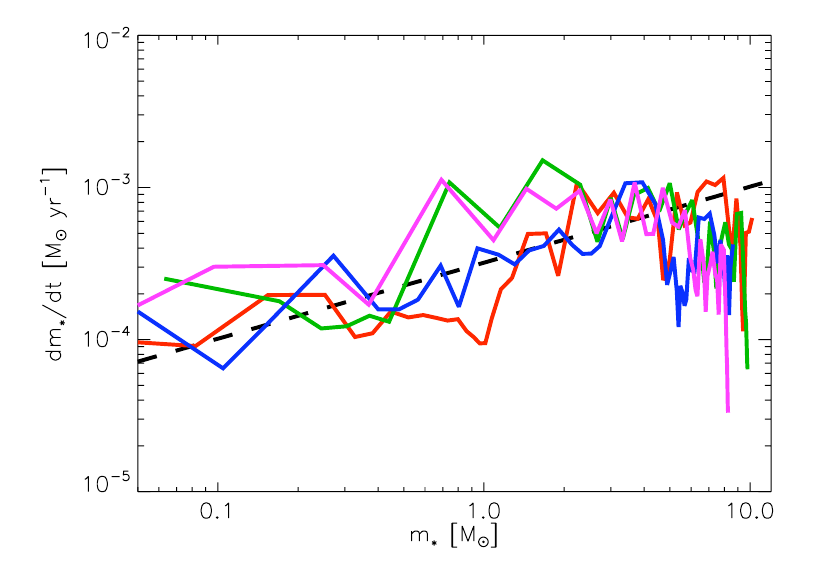}
\caption{\label{fig:hmaccrate}
Accretion rate versus stellar mass for the four most massive stars present at the end of run TuW. Colored unbroken lines indicate the measured simulation accretion rates, while the dashed black line is the prediction of the \citet{mckee03a} model (Equation \ref{eq:mtaccrate}). The simulation accretion rates have been smoothed over 500 yr timescales to reduce scatter.
}
\end{figure}

Run TuW is the first published simulation that includes radiative and protostellar outflow feedback, produces an IMF that is in good agreement with the observed IMF over a broad mass range, and forms a large enough cluster for there to be massive stars present. It is therefore important to pay particular attention to the processes by which those massive stars form. We turn now to the properties of the massive stars in run TuW.

There has been considerable discussion in the literature about whether massive stars form from distinct massive protostellar cores \citep{padoan95a, padoan02a, mckee02a, mckee03a, krumholz07a, hennebelle08a, hennebelle09a}, or whether all stars are born from cores with masses $\la 1$ $\msun$, and massive stars subsequently grow from these small seeds by Bondi-Hoyle accretion \citep{bonnell97a, bonnell01a, bonnell01b, bonnell04a, bonnell06d, bonnell02a, bonnell06c, bate05a, smith09b, smith09a}. A number of authors have also proposed hybrid models, in which massive stars form from gravitationally bound gas structures, but these structures are assembled and fed from larger scales at the same time as they form massive stars \citep{peretto06a, wang10a}. To address this question, we examine the four most massive stars present at the end of run TuW; these have masses of $10.8$, $9.8$, $8.8$, and $8.3$ $\msun$, respectively, and thus each is large enough that, even if it were to accrete no further, it would be expected to end its life as a supernova. For comparison, we also examine the four stars whose masses are closest to the median mass at the end of the simulation, $0.34$ $\msun$. For each of these stars, we identify the point in space and time at which that star first appeared in our simulations, and examine the gas density distribution in its vicinity.

We show the results in Figure \ref{fig:snapshots_hmcores} for the high mass cores and Figure \ref{fig:snapshots_lmcores} for the low mass cores. To facilitate comparison with observations, in addition to showing the true gas density distribution, we show the distribution smeared with a $1700$ AU Gaussian beam; we choose this size scale because it is approximately the spatial resolution of the highest published resolution maps of massive cores \citep[e.g.][]{beuther04b, bontemps10a}, though the Atacama Large Millimeter Array (ALMA) will soon produce images at significantly higher resolution. Figure \ref{fig:snapshots_hmcores} demonstrates that the massive stars in our simulation form in distinct, massive overdensities that can be identified as cores. Their characteristic sizes, determined from visual inspection, are roughly $0.01$ pc. Comparing the gravitational and kinetic energies in this structures shows that they are roughly gravitationally bound and virialized. The flows within them are highly supersonic, producing a filamentary morphology. Nonetheless, these objects are not highly sub-fragmented. There are at most one or two density maxima in each one, not many density maxima. These structures look much like the turbulent cores posited in the \citet{mckee03a} theory for massive star formation. When smeared on a resolution of 1700 AU, distinct centrally-condensed structures remain visible for three of the four massive stars, indicating that these objects would be detectable as massive cores in an observation. 


\red{It is important to understand that our analysis says nothing about the Lagrangian trajectories of the fluid elements that eventually coalesce to form the massive stars in our simulations, a topic that has previously received extensive investigation by \citet{bonnell04a} and \citet{smith09b, smith09a}, among others. It may well be that particular fluid elements that are present in the cores at the time shown in Figure \ref{fig:snapshots_hmcores} do not accrete onto the final star and are instead accreted by other stars or torn off by turbulent motions, while fluid elements not present in the core at the time shown are eventually accreted into the final star. Indeed, \citet{mckee03a} predicted in their analytic model that turbulent cores should over the course of their lives interact with a surrounding gas mass comparable to that which eventually ends up in their central stars. However, the fact that the {\it Lagrangian} elements making up a core change with time is irrelevant to the question of whether, as a massive star forms, it sits at the center of a gravitationally bound {\it Eulerian} structure. Figure \ref{fig:snapshots_hmcores} shows that it does.}

We can make the link between the massive cores and the stars they form more quantitative by comparing to the massive core evolution model of  \citet{mckee02a, mckee03a} and \citet{tan04a}. This model predicts that the accretion rate onto a star as a function of its mass should be
\begin{equation}
\label{eq:mtaccrate}
\dot{m}_* = 1.2\times 10^{-3} \left(\frac{m_{*f}}{30\,\msun}\right)^{3/4} \Sigma_{\rm cl}^{3/4} \left(\frac{m_*}{m_{*f}}\right)^{1/2}\;\msun\mbox{ yr}^{-1},
\end{equation}
where $m_*$ is the star's instantaneous mass, $m_{*f}$ is its final mass, $\Sigma_{\rm cl}$ is the surface density of the molecular clump from which it forms, and we have used \citeauthor{mckee03a}'s fiducial parameter choices, with the exception that we have increased the accretion rate by a factor of 2.6 to include subsonic contraction, following \citet{tan04a}. To evaluate this equation and compare it to our simulations, we take $m_{*f} \approx 10$ $\msun$, since this is roughly the mass of our four most massive stars at the end of the simulation. For $\Sigma_{\rm cl}$, we note that, in the simulation, the core is better-defined than the clump, so we adopt \citeauthor{mckee02a}'s result with $\Sigma_{\rm cl}$ replaced by $\Sigma_{\rm core}$. In their fiducial model these agree to within a factor $\simeq 1.2$, so this does not significantly affect the accretion rate. As shown in Figure \ref{fig:snapshots_hmcores}, our cores have masses of order $10$ $\msun$ in radii of order $0.01$ pc, which corresponds to $\Sigma_{\rm cl} = 6.6$ g cm$^{-2}$. With these parameter choices, in Figure \ref{fig:hmaccrate} we plot the accretion rate as a function of stellar mass for the four most massive stars at the end of the simulation, whose cores are shown in Figure \ref{fig:snapshots_hmcores}, and compare to the \citeauthor{mckee03a} prediction. As the plot shows, the simulation accretion rates agree quite well with the analytic predictions.

In contrast, the cores that give rise to low mass cores (Figure \ref{fig:snapshots_lmcores}) are quite noticeably different from the high mass ones. In three of the four cases (the first, third, and fourth columns in the Figure) they are also centrally-condensed lumps of gas.
However, unlike the massive cores they are highly sub-fragmented and show many density maxima. Clearly these objects are not single cores, but instead tightly-packed agglomerations of many smaller cores. For the final low mass core (shown in the second column of Figure \ref{fig:snapshots_lmcores}) the point at which the star forms is a slight overdensity in the middle of a filament, and there is no centrally-concentrated object at all. Thus massive cores and low mass cores have clearly distinct properties. However, we also find that these differences are completely indistinguishable in the smeared images, indicating that it is not possible to distinguish true high mass cores from agglomerations of low mass ones with the resolution available in pre-ALMA telescopes, at least for objects at the $\sim$kpc distances typical of massive star-forming regions. This conclusion is consistent with that of \citet{offner12a}.

\red{It is important to note that the differences between high and low mass stars is not simply a function of formation time. It is certainly true that the most massive stars at the end of the simulation preferentially began forming early. However, their greater masses are far less a reflection of this than it is of their different formation environments. The four massive stars grow at time-averaged rates of $3.6 - 4.6\times 10^{-4}$ $\msun$ yr$^{-1}$, compared to $1.7 - 8.8\times 10^{-5}$ $\msun$ yr$^{-1}$ for the low mass stars. At the accretion rates typical of the low mass stars, it would require $\sim 10\tff$ for one of them to grow to the $\sim 10$ $\msun$ typical of the massive stars. The massive stars are not simply those that form first; they are those that form surrounded by coherent, bound, non-subfragmented structures that provide high accretion rates.} This is somewhat similar to the competitive accretion model in that massive stars' preferred locations at the centers of collapsing regions that provides their high accretion rates. However, it differs from competitive accretion in that these cores are non-subgfragmented and have masses at the same order of magnitude as the final stars, $\sim 10$ $\msun$, and therefore intermediate between that of the entire star cluster, $\sim 10^3$ $\msun$ and the thermal Jeans mass, $\sim 1$ $\msun$. In the competitive accretion model such structures should be absent, because everything fragments down to the thermal Jeans mass \citep{bonnell04a, bate05a} and some objects subsequently grow to larger masses by Bondi-Hoyle accretion. There are no $\sim 10$ $\msun$ objects that do not subfragment in competitive accretion.

Finally, we note that both the high mass and the low mass cores are above the column density threshold $\Sigma > 1$ g cm$^{-2}$ for massive star formation posited analytically by \citet{krumholz08a} and confirmed numerically by \citet{krumholz10a}. This means that both the high mass and low mass cores have the {\it potential} to form massive stars; indeed, in one of the four cases shown in Figure \ref{fig:snapshots_lmcores}, the low mass star is in fact forming in a core that puts most of its mass into a single high mass star. That does not appear to be the case for the other three low mass stars shown in the Figure, however. Thus a high column density is clearly a necessary but not a sufficient condition for massive star formation. A high column density allows radiative heating to suppress the growth of gravitational instabilities that would lead to fragmentation and prevent a massive star from forming. However, if the turbulent density field present before a star begins radiating is already highly non-linearly fragmented, as is the case for several of the low mass cores shown in Figure \ref{fig:snapshots_lmcores}, radiative heating will not undo this fragmentation and prevent the core from forming a small cluster of low mass stars rather than a few massive ones.

\subsection{Stellar Multiples}

\begin{figure}
\epsscale{1.1}
\plotone{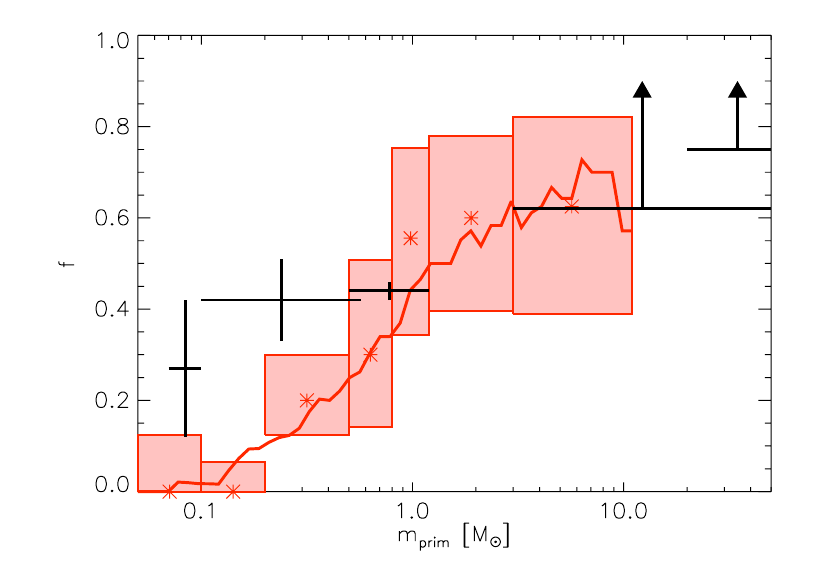}
\caption{
\label{fig:binstat}
Multiplicity fraction $f$ as a function of system primary mass $m_{\rm prim}$ in run TuW. The thick red line shows $f$ as a running average. The light red boxes show $f$ computed over discrete bins in $m_{\rm prim}$. In each case, the width of the box shows the primary mass range for that bin, the asterisk shows the mean multiplicity fraction for stars in that bin, and the vertical extent of the box shows the statistical uncertainty on that value, computed as described in the text. Finally, black crosses indicate observational results, with the horizontal width indicating the mass range for the observations and the vertical range showing the stated uncertainty. The two highest mass observational data points are lower limits, indicated by the upward arrows. The data shown are taken from, from left to right, \citet{basri06a} and \citet{allen07b} (shown as a single combined point), \citet{fischer92a}, \citet{raghavan10a}, \citet{preibisch99a}, and \citet{mason09a}; the data compilation shown here is the same as that in \citet{bate12a}.
}
\end{figure}

\begin{figure}
\epsscale{1.1}
\plotone{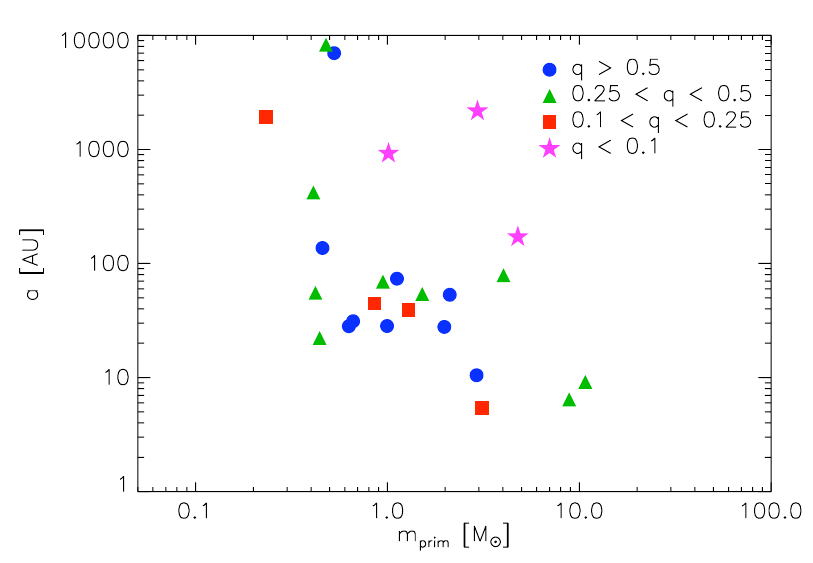}
\epsscale{1.0}
\caption{
\label{fig:binprop}
Semi-major axis versus primary star mass for all the binaries in our simulations. For triple and quadruple systems, we plot them only once, showing the properties of the most bound pair of stars. Points are coded by the mass ratio of the system: purple stars for $q<0.1$, red squares for $q=  0.1-0.25$, green triangles for $q = 0.25 - 0.5$, and blue circles for $q > 0.5$.
}
\end{figure}

It is also illuminating to consider the properties of the stellar multiples that form in run TuW, since producing the correct multiplicity fraction has been proposed as test for star formation models in addition to producing the correct IMF \citep[e.g.][]{bonnell07a}. We therefore examine the final time slice for this run.\footnote{In this section of the paper alone we do not exclude stars smaller than $0.05$ $\msun$ from consideration, but we consider them only as companions to larger stars. We allow them to count in this capacity because to omit them would artificially make it impossible for stars near our $0.05$ $\msun$ cut to have companions.}  Extracting the fraction of stars in multiple systems from the simulation requires some care, as pointed out by \citet{bate09b}. Many of the stars in our simulation form a bound cluster, and thus many stars are bound to many other stars, often in hierarchical structures consisting of dozens of individual stars; for example a binary and a triple system may orbit one another, and these in turn may have additional stars orbiting them. Such agglomerations would be extremely unlikely to survive dynamically even for the lifetime of a massive star, and would break up if we could continue the simulation further.

Thus we follow \citeauthor{bate09b} in defining stellar multiplicity via the following algorithm. We first compute the total energy (gravitational plus kinetic in the center of mass frame) pairwise for each pair of stars in the simulation. We find the most bound system and replace it with a single point mass, with a mass equal to the sum of the two components, a position located at their center of mass, and a momentum equal to the sum of their two momenta. We then continually repeat this process, with the exception that we do not create aggregates consisting of more than four individual stars; should the most bound system contain five our more stars, we proceed to the next most bound pair with fewer than five members instead.\footnote{The results are not particularly sensitive to the choice of four as the maximum size of a system, as long as we stop at some point well short of allowing the entire cluster to be considered a single large star system.} We terminate the process once there are no more bound pairs consisting of fewer than five individual stars. At the end we are left with a list of star systems, some single and some containing up to four individual stars.

Given this list, we can compute the fraction of multiple systems as a function of primary star mass. For a set of star systems, we define the multiplicity fraction
\begin{equation}
f = \frac{B+T+Q}{S+B+T+Q},
\end{equation}
where $S$, $B$, $T$, and $Q$ and the numbers of single, binary, triple, and quadruple systems, respectively.\footnote{Following \citet{bate09b} and \citet{hubber05a}, we measure this quantity rather than either the companion star fraction $(B+2T+3Q)/(S+B+T+Q)$ or the fraction of stars in multiple systems $(2B+3T+4Q)/(S+2B+3T+4Q)$ because it is more robustly determined observationally. If a new member of a multiple system is found, for example leading to a binary being reclassified as a triple, $f$ does not change, while the companion star fraction and the fraction of stars in multiple systems does.} We choose our sets of star systems in two ways. The first is as a running average; for a primary mass $m_{\rm prim}$, we compute $f$ considering all systems for which the primary mass is within half a dex of $m_{\rm prim}$. The second is in discrete bins, chosen to roughly match the mass ranges selected in observational surveys. We consider primary mass bins in the range $0.05 - 0.1$ $\msun$, $0.1-0.2$ $\msun$, $0.2 - 0.5$ $\msun$, $0.5-0.8$ $\msun$, $0.8 - 1.2$ $\msun$, $1.2 - 3$ $\msun$, and $>3$ $\msun$. In addition to the mean value, we compute the statistical uncertainty in this value for each bin.\footnote{We determine the statistical uncertainty by assuming that there is a true multiplicity fraction $f_{\rm true}$ for stars in that bin, and that our sample of systems in that bin represent a series of random drawings that follow a binomial distribution. From these assumptions, we can compute the probability distribution for $f_{\rm true}$ given the measured multiplicity fraction in the simulations $f_{\rm sim}$ and the number $N_{\rm sys}$ of systems in that mass bin. We compute the uncertainty by finding the range of values for $f_{\rm true}$ that enclose the central $68\%$ of the probability distribution. Note that this range is not in general symmetric about $f_{\rm sim}$.}

We plot the results in Figure \ref{fig:binstat}. We see that the simulations generally agree well with the observational constraints, with the multiplicity fraction reaching near unity for stars larger than a few $\msun$, and declining to below $0.5$ for stars smaller than $\sim 0.5$ $\msun$. Our simulations somewhat underproduce binaries at the lowest masses, which is likely a resolution effect, arising because low mass binaries must be very close in order to remain bound, and our simulation resolution makes this difficult. In our simulation we soften particle-particle gravity forces on a scale of 0.25 cells, or $5.8$ AU, and gas-particle gravity forces are necessarily smoothed on the grid scale of 23 AU. Thus we cannot easily make binaries tighter than $\sim 10$ AU. At this distance the Keplerian speed around a 0.05 $\msun$ object is only 2 km s$^{-1}$, comparable to the velocity dispersion in the cluster. Thus low mass binaries formed in our simulation will tend to be broadened and disrupted, and we cannot resolve the tighter binaries that will tend to be hardened. This leads to an artificial reduction in the binary fraction at low masses, a phenomenon also noted by \citet{bate12a}.

In Figure \ref{fig:binprop} we illustrate some of the properties of our multiple systems. Systems with more massive primaries tend to have the smallest separations, as a result of dynamical hardening and, in some cases, of a companion having been born in the disk of the primary. The companions to the most massive stars also tend to be fairly massive, with mass ratios of $0.25 - 0.5$; this is inconsistent with their having been drawn randomly from the IMF. These are often triple or quadruple systems. Thus we see that the massive stars in our simulation tend to form Trapezium-like structures. In contrast, at near-Solar masses, the range of semi-major axes and mass ratios is extremely broad.

\subsection{Accretion Variability and Outbursts}

\begin{figure*}
\epsscale{1.1}
\plotone{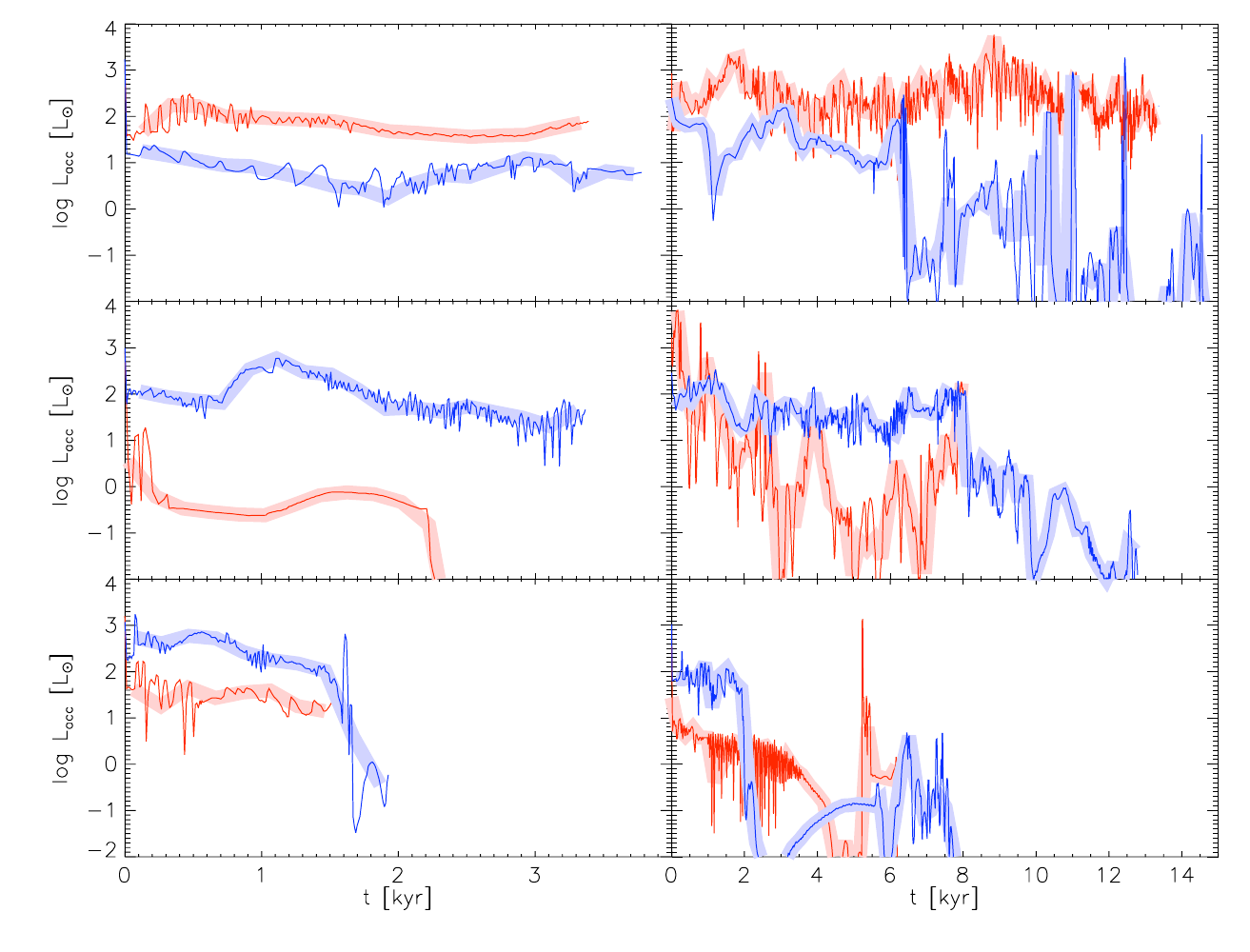}
\caption{\label{fig:accvar}
Accretion rate versus time for a sample of 12 randomly-selected stars in run TuW. Thick pale lines show accretion luminosities averaged over 200 yr timescales, while thin darker lines show the accretion luminosity computed over $10-20$ yr timescsles, the finest available given the frequency with which we output simulation data. There is no distinction between red and blue curves; we simply use two different colors to make the two stellar accretion histories shown in each panel more easily distinguishable. Note that the time axes are different for the left and right sides. All times are relative to the instant when a star first appears in the simulations, and plots continue to the end of the simulation.\\
}
\end{figure*}

A final useful datum to be extract from run TuW is the amount of accretion variability for low mass stars, which is of interest for its relevance to the protostellar luminosity problem and the origin of FU Ori outbursts, although due to the duration of our simulation we can only address these issues as they apply to class 0 and I objects, not class II sources. To characterize the degree of luminosity variability we find, we select 12 stars at random at the end of our simulation. The final masses of these stars range from $0.055$ to $1.9$ $\msun$. For each star we measure its accretion luminosity, which in our code is taken to be
\begin{equation}
L_{\rm acc} = 0.75 \frac{G m_* \dot{m}_*}{r_*},
\end{equation}
where $m_*$, $\dot{m}_*$, and $r_*$ are the star's instantaneous mass, accretion rate, and radius, and the factor of $0.75$ accounts for the energy used to drive protostellar outflows \citep[for details see the Appendix of][]{offner09a}. Our simulation outputs are spaced roughly $10-15$ yr apart, so this represents the minimum timescale on which we can study variability. Outputs are 80 fine grid time steps apart, so this timescale is numerically well resolved.

In Figure \ref{fig:accvar} we show the accretion history of each of our 12 stars, both at the maximum temporal resolution of the simulations, and smoothed over 200 yr timescales. In the figure, the zero of time is the point at which a given star forms, and we plot the accretion history for the remainder of the simulation. The figure demonstrates several interesting results. First, the majority of stars have relatively smooth luminosity histories when averaged over 200 yr timescales. For only a few examples do order of magnitude variations in the luminosity occur on less than timescales of several kyr. The variability is somewhat larger when measured at the maximum temporal resolution of the simulation, but for most stars this is not a large effect. However, there are three exceptions: the star shown in blue in the upper right panel, the star shown in blue in the lower left panel, and the star shown in red in the lower right panel. All of these stars experience sudden increases in luminosity on timescales below our ability to resolve given the frequency of our output. During these spikes, the luminosity rises by $1-2$ dex compared to the long-term average. These are plausibly FU Ori-type outbursts, although we caution again that these outbursts are occurring in class I sources, not true T Tauri stars.

\section{Discussion and Conclusion}
\label{sec:discussion}

\subsection{The Role of Protostellar Outflow Feedback}

In our simulations, we find that protostellar outflow feedback is not particularly effective. Including outflows reduces the star formation rate by only $\sim 20\%$, comparable to the mass fraction that is ejected from young stars by out subgrid model for protostellar winds. This result at first appears to contradict those of previous studies, including \citet{li06b}, \citet{nakamura07a}, \citet{matzner07a}, \citet{wang10a}, and \citet{cunningham11a}, all of whom find that outflow feedback is important. Our results also contradict those of \citet{hansen12a}, who find that outflow feedback greatly reduces the efficacy of radiative feedback, because it reduces the accretion rate and thus the protostellar luminosity.

Some of our differences from previous results are a function of what effects are included in our simulations, and in previous work. \citet{wang10a} note that outflow feedback is only effective as a long-term driver of turbulence in the presence of magnetic fields. Fields facilitate transfer of momentum between gas parcels, while in their absence most of the momentum injected into a protocluster by outflows is simply lost, as the outflows break out of the cloud and deposit their momentum and energy outside its boundaries. Since our simulations lack magnetic fields, we likely suffer from a similar underestimate of outflow efficacy.

A second source of difference is likely to be our choice of parameters. We have simulated a fairly massive, high surface density cloud representative of the typical Galactic star-forming region, but with properties quite distinct from those of the star-forming regions closest to the Sun. All of the previous simulations mentioned above have chosen properties typical of these lower density regions. Analytic models suggest that protostellar winds are only able to eject significant mass from clusters with escape velocities below $v_{\rm esc} \sim 7$ km s$^{-1}$ \citep{matzner00a, matzner07a}. In run TuW, the escape velocity is $v_{\rm esc} \approx \sqrt{G M_c/(\ell_c/2)} = 4.3$ km s$^{-1}$, within a factor of 2 of the analytic estimate. The comparable figure for the cluster simulated by \citet{wang10a}, which is modeled after NGC 1333, is a factor of two lower: $v_{\rm esc} \approx 2.6$ km s$^{-1}$. For \citet{hansen12a}, who adopt initial conditions modeled after $\rho$ Ophiuchus, it is $v_{\rm esc} \approx 1.6$ km s$^{-1}$. Thus it is not surprising that outflows should be much more effective in those simulations than in run TuW. Indeed, placing the cluster masses and surface densities of these simulations on \citet{fall10a}'s diagnostic diagram for where different sorts of feedback are effective (their Figure 2) immediately predicts this dichotomy.

Given that our simulations likely differ from previous work on the importance of outflow feedback due to both a physical deficiency (lack of magnetic fields) and a choice of parameters that is closer to the typical region than most previous work, it is hard to draw general conclusions about the importance of outflow feedback in regulating star formation. Resolution of this question will have to await future magnetohydrodynamic simulations that probe the higher density regime we have explored in this work.

\subsection{Implications for Massive Star Formation}

The picture of massive star formation that emerges from our simulations is 
\red{generally}
consistent with the turbulent core model proposed by \citet{mckee02a, mckee03a}. 
\red{The massive stars form at the centers of well-defined, turbulent, centrally-concentrated structures, and these structures feed mass onto them at a rate that is consistent with the predictions of the \citeauthor{mckee03a} model.}
In contrast, the regions from which low mass stars form are quite noticeably different. They are either messy regions consisting of many small density peaks and no clear central concentration, or they are small regions of filaments. Thus the basic core to star mapping proposed in the turbulent core model appears to describe our simulation fairly well.

However, we also do see elements of the alternative competitive accretion model \citep[and references therein]{bonnell07a} operating as well. In particular, our massive stars do all form as part of small sub-clusters and experience significant dynamical interactions. These interactions appear to be important in shaping the multiplicity properties of the resulting stars, and in producing the Trapezium-like systems in which most of our massive stars find themselves at the end of the simulation.

\subsection{Implications for the IMF and the Star Formation Rate}
\label{sec:implications}

Run TuW represents the first simulation published to date that reproduces the observed IMF in a cluster like the ONC that is large enough to contain massive stars, and where the peak of the mass function is determined by a fully self-consistent calculation of gas thermodynamics. Previous simulations that have had success in reproducing the IMF have either examined small, low-density star-forming regions that would not be expected to produce massive stars \citep{offner09a, bate09a, bate12a, urban10a, hansen12a}, or have relied on a parameterized, non-self-consistent equation of state to determine the location of the IMF peak \citep[e.g.][]{bate05a, jappsen05a}.

The success of run TuW, in contrast to the failures of runs SmNW and TuNW, suggests that obtaining the correct IMF from a self-consistent simulation of a typical star-forming environment is not as simple as some authors have posited \citep[e.g.][]{bonnell07a}. While a lognormal function with a powerlaw tail at high masses appears to be a fairly generic result regardless of what physics is included in the simulations, getting the peak of the lognormal to lie at the correct position in a calculation where it is determined self-consistently rather than through a hand-imposed equation of state requires careful attention to the thermodynamics of the gas, which is in turn determined primarily by the accretion luminosity of protostars.

This requires several ingredients to work correctly. As conjectured in Paper I, the star formation rate cannot be too high, \red{and star formation cannot become too centrally cocentrated}; if it is, the resulting accretion luminosity becomes so high that formation of low mass stars is suppressed and the IMF peak marches to ever-increasing mass with time. In addition, outflows appear to make a small but significant contribution by both reducing the masses of individual accreting protostars, reducing the accretion luminosity, and ejecting mass from the warm regions near accreting stars, increasing fragmentation. The combination of these effects shifts the mean mass downward by a factor of $\sim 2$. Our results suggest that future simulations of gas collapse and fragmentation, if they are to reproduce the observed IMF while treating the gas thermodynamics self-consistently, must at a minimum include these ingredients.

We obtain a
\red{reduction in the rate and degree of concentration of star formation}
in run TuW mainly because we have
\red{starting our simulations with a self-consistent density and velocity field and by}
embedding the protocluster gas clump in a realistic surrounding turbulent molecular cloud that continually feeds energy into it via a turbulent cascade, rather than simulating a \red{smooth,} isolated clump as in most previous work. However, the star formation rate per free-fall time we obtain is still an order of magnitude too high compared to observations, likely as a result of other mechanisms we have not included. For example, \citet{price09a} and \citet{padoan11a} both find that magnetic fields with strengths comparable to those in observed molecular clouds reduce star formation rates by a factor of a few to $\sim 10$. Depending on protocluster properties, ionized gas and radiation pressure may also contribute to reducing the star formation rate \citep[e.g.][]{fall10a, lopez11a}. Ultimately, since accretion luminosity plays a critical role in regulating the IMF, the problems of determining the star formation rate and the IMF cannot be fully separated. A truly  accurate simulation must reproduce both.

\acknowledgements This work was supported by an Alfred P.\ Sloan Fellowship (MRK); the NSF through grants CAREER-0955300 (MRK) and AST-0908553 (CFM and RIK); NASA through ATFP grant NNX09AK31G (RIK, CFM, and MRK) and a Chandra Space Telescope grant (MRK); and the US Department of Energy at LLNL under contrast DE-AC52-07NA (RIK). Support for computer simulations was provided by an LRAC grant from the NSF through Teragrid resources and NASA through grants from the ATFP and Chandra Programs.


\end{document}